\documentclass[12pt]{article}

\usepackage{jheppub}
\makeatletter
\def\@fpheader{\relax}
\makeatother

\usepackage{amsthm,amsmath,amssymb,mathtools}
\usepackage{mathrsfs}
\usepackage{bm}
\usepackage{bbm}
\usepackage{graphicx}
\usepackage{float}
\usepackage{caption}
\usepackage{subcaption}
\usepackage[most]{tcolorbox}
\usepackage[mathscr]{euscript}
\usepackage{dsfont}
\usepackage[mathscr]{euscript}
\usepackage{mathrsfs}
\usepackage{hyperref}
\usepackage{tikz}

\newcommand\blfootnote[1]{%
	\begingroup 
	\renewcommand\thefootnote{}\footnote{#1}%
	\addtocounter{footnote}{-1}%
	\endgroup 
}
\def\mD{\mathcal{D}}

\def\mA{\mathcal{A}}
\def\mN{\mathcal{N}}
\def\mL{\mathcal{L}}

\def\mW{\mathcal{W}}

\def\mba{\mathbf{a}}
\def\mbb{\mathbf{b}}

\newenvironment{nohyphens}{%
	\hyphenpenalty=10000
	\exhyphenpenalty=10000
	\sloppy %
}{\par}

\title{Notes on su$(1,2)\oplus$u$(1)$ Chern-Simons theory and Torsional Newton-Cartan gravity}

\author[a]{Yang Lei}
\author[a]{\!\!, Dong Zhang \blfootnote{*The authors are ordered purely alphabetically and should all be viewed as the co-first authors. }}

\affiliation[\,a]{Institute for Advanced Study \& School of Physical Science and Technology,  \\ Soochow University, Suzhou 215006, P.R.~China}

\emailAdd{leiyang@suda.edu.cn}
\emailAdd{zdgygy@outlook.com}

\abstract{\begin{nohyphens}
In this study, we investigate three-dimensional torsional Newton-Cartan (TNC) gravity by gauging the su$(1,2)\oplus$u$(1)$ algebra and construct its action using the Chern-Simons theory. 
This TNC exhibits novel features, including the fact that the gauge fields associated with both dilatation and rotation symmetries transform non-trivially under Galilean boosts. 
This theory also reproduces the Schr\"odinger gravity acquired by gauging the extended $z=2$ Schr\"odinger algebra via a large speed of light ($1/c$)-expansion. 
In particular, we explain that the $z=2$ Lifshitz vacuum solution appearing in Schr\"odinger gravity is related to the null reduction of 4d $\Omega$-background up to a conformal factor. 
Based on these results, we revisit the identification between the extended Schr\"odinger algebra and bosonic analogue of super BMS algebra.
We interpret that this relation originates from the $\mathcal{W}_3^{(2)}$ algebra which acts as the bosonic analogue of $\mathcal{N}=2$ superconformal algebra. 
	\end{nohyphens}
}

\date{}

\begin{document} 
	
	\maketitle
	
	\section{Introduction}\label{sec:intro}

Significant progress has been made in understanding three-dimensional gravity since its equivalence with the Chern-Simons theory was established \cite{Witten:1988hc}. 
Various gravitational theories can be constructed by selecting appropriate gauge algebras for the Chern-Simons gauge field. 
For instance, the Chern-Simons theory with the algebra so$(2,2)=$sl$(2,\mathbb{R}) \oplus $sl$(2,\mathbb{R}) $ corresponds to AdS$_3$ gravity, whereas the theory with iso$(1, 3)$ corresponds to the flat limit of gravity \cite{Brown:1986nw}.
Over the past decade, this equivalence has been instrumental in developing gravitational theories that extend beyond Einstein's gravity. A notable example of such an extension is higher spin gravity, which is based on the algebra hs$[\lambda] \oplus $hs$[\lambda]$ (or sl$(N,\mathbb{R}) \oplus$ sl$(N,\mathbb{R})$ for finite truncation) \cite{Campoleoni:2010zq, Henneaux:2010xg,Bergshoeff:1989ns, Gaberdiel:2010pz, Gaberdiel:2012uj,Gutperle:2011kf}. 
See also \cite{Campoleoni:2024ced,Kiran:2014dfa,Ammon:2012wc} for reviews on this topic. 

Interestingly, studies \cite{Afshar:2012nk,Gary:2012ms,Gary:2014mca} and subsequent work \cite{Gutperle:2013oxa,Gutperle:2014aja,Beccaria:2015iwa} have demonstrated the existence of Lifshitz \cite{Kachru:2008yh} and Schr\"odinger geometries \cite{Son:2008ye} characterized by a dynamical exponent $z=N-1$ where $N$ denotes the rank of the sl$(N,\mathbb{R})$ algebra of the Chern-Simons gauge theory \cite{Hartnoll:2009sz}.
These geometries have been extensively studied in the context of non-relativistic holography, which seeks to establish a duality between gravitational theories and strongly coupled non-relativistic field theories. 
Such holographic frameworks have proven useful for understanding condensed matter systems, particularly those exhibiting critical behavior or non-Fermi liquid phases, where traditional perturbative methods often fail. 
See \cite{Prohazka:2017lqb,Taylor:2015glc, Horava:2009uw, Hartnoll:2009sz,Taylor:2008tg} for a non-exhaustive list of reviews on these topics.
The Lifshitz and Schr\"odinger solutions discovered in the Chern-Simons theory provide concrete gravitational duals for these systems, offering insights into the dynamics of strongly correlated electrons, quantum critical points, and other phenomena in non-relativistic settings. 
However, as clarified in \cite{Lei:2015ika, Hartong:2016yrf}, the $z=2$ Lifshitz solutions identified in \cite{Afshar:2012nk,Gary:2012ms,Gary:2014mca} are more accurately interpreted as solutions within the framework of Chern-Simons Newton-Cartan gravity.
This connection positions the Chern-Simons theory as a potential bridge between Newton-Cartan type gravitational theories and condensed matter systems, offering a useful framework for understanding strongly coupled non-relativistic field theories.

Newton-Cartan gravity, which plays a central role in this context, can be categorized into three distinct types \cite{Hartong:2014oma}: (i) torsionless $d\tau=0$ \cite{Andringa:2010it}, (ii) twistless torsional (TTNC) with $\tau\wedge d\tau =0$ and (iii) torsional (TNC) \cite{Bergshoeff:2014uea,Hartong:2015zia} (no conditions on $\tau$). 
The exploration of the three-dimensional Chern-Simons Newton-Cartan theory can be traced back to \cite{Papageorgiou:2009zc}, where torsionless Newton-Cartan gravity was constructed by gauging the extended Bargmann algebra. 
Building on this foundation, \cite{Bergshoeff:2016lwr} developed Newton-Cartan supergravity by gauging the extended super Bargmann algebra, while \cite{Hartong:2016yrf} demonstrated that twistless torsional Newton-Cartan gravity can be constructed by gauging the extended Schr\"odinger algebra. 
Further advancements in the study of the Chern-Simons Newton-Cartan gravity are documented in \cite{Papageorgiou:2010ud,Hartong:2017bwq,Chernyavsky:2019hyp,Gomis:2019nih,Aviles:2018jzw,Ozdemir:2019tby,Concha:2019mxx,Concha:2022muu,Concha:2020eam}. 
Despite these achievements, two significant questions remain. First, it is still unclear how to construct the most general type (iii) torsional Newton-Cartan gravity within the Chern-Simons framework, which requires the clock one-form to satisfy the condition $d\tau -z b\wedge \tau \neq 0$. 
Second, although general twistless torsional Newton-Cartan (TTNC) gravity permits the dynamical exponent $z>1$, only $z=2$ TTNC theory was found in the Chern-Simons theory \cite{Hartong:2016yrf}.

In this paper, we study the Chern-Simons theory with gauge algebra su$(1,2)\oplus$u$(1)$. 
As we demonstrate, the resulting theory is equivalent to torsional Newton-Cartan gravity, which was also explored in \cite{Baiguera:2024vlj}. 
This study was motivated by two key questions in this field.
\begin{itemize} 
	\item Understand the broader relevance of su$(1,n)$-type symmetries in non-relativistic holography and condensed matter systems
	\item Identify the ``seed theory" underlying twistless torsional Newton-Cartan (TTNC) gravity \cite{Hartong:2016yrf}.
\end{itemize}
We will explain them below.

Recent studies have revealed that su$(1,n)\oplus$u$(1)$-type symmetries arise naturally in the null reduction limit of a Minkowski background with an $\Omega$ deformation \cite{Lambert:2021nol, Lambert:2020scy,Lambert:2021nol,Smith:2023jjb, Baiguera:2024vlj,Lambert:2024uue,Lambert:2024yjk}. 
This  algebra can be interpreted as the conformal completion of the $z=2$ Lifshitz algebra  \cite{Lambert:2021nol,Baiguera:2024vlj},
making models with 
su$(1,n)$ global symmetry compelling candidates for investigating non-relativistic holography with conformal symmetry.
Another class of non-relativistic models exhibiting su$(1,n)$-type symmetry is known as the Spin Matrix theory \cite{Harmark:2007px,Harmark:2014mpa, Harmark:2019zkn,Harmark:2020vll, Baiguera:2020jgy, Baiguera:2020mgk, Baiguera:2021hky,Baiguera:2022pll,Baiguera:2023fus,Baiguera:2024vlj}.
These models emerge from the non-relativistic limit of $\mN=4$ SYM. 
The Spin Matrix theory with the largest spin group possesses a PSU$(1,2|3)$ global symmetry and is believed to have a ground state dual to dual to $1/16$-BPS black holes in AdS$_5$ \cite{Kinney:2005ej,Kunduri:2006ek,Choi:2018hmj,Cabo-Bizet:2018ehj,Benini:2018ywd}.  
However, despite the success in SU$(2)$ subsector \cite{Harmark:2016cjq}, the general understanding of the dual gravitational theories remains in its early stages. 
By studying the structure of the Chern-Simons theory with su$(1,2)\oplus$u$(1)$ symmetry, we aim to uncover the features of gravitational theories with similar symmetries, potentially improving our understanding of the gravitational dual of non-relativistic conformal field theories.

\begin{figure}
	\centering
	\includegraphics[trim=4cm 22cm 7cm 4cm,width=0.7\linewidth]{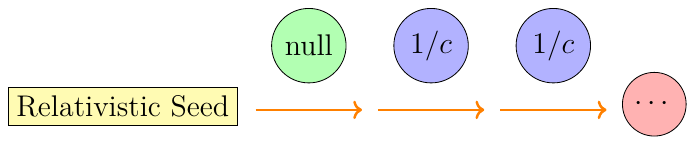}
	\caption{Schematically, a relativistic Lorentzian manifold with a null direction can be null reduced to the Newton-Cartan manifold. The geometry can be further acted by $1/c$ to any order to generate novel Newton-Cartan gravities. 
More DLCQ transformations could be allowed if one perform duality transformations on the non-Lorentzian manifold \cite{Blair:2024aqz,Gomis:2023eav}. }
	\label{fig:seed}
\end{figure}

The second aspect is related to a central challenge in identifying the seed theory that underlies the Chern-Simons TTNC gravity \cite{Hartong:2016yrf}.
By ``seed theory," we refer to the initial gravitational framework (e.g., Einstein gravity) that, when subjected to universal algorithms such as null reduction or $(1/c)$-expansion, generates new non-Lorentzian theories at subleading orders. 
See figure \ref{fig:seed}.
The null reduction algorithm, widely used to derive the Newton-Cartan geometry from a Lorentzian geometry with a null direction \cite{Duval:1984cj,Julia:1994bs, Christensen:2013lma,Christensen:2013rfa,Harmark:2018cdl}, and the large speed of light ($1/c$)-expansion \cite{Dautcourt:1996pm,VandenBleeken:2017rij, Harmark:2017rpg, Hansen:2019pkl,Hansen:2019svu,Hansen:2020pqs, Hansen:2020wqw}, recently reviewed in \cite{Hartong:2022lsy}, are two systematic approaches for generating Newton-Cartan gravity. 
The $(1/c)$-expansion generates, at each order, a hierarchy of gravitational actions supplemented by novel fields arising from higher-order expansions of fields, whose transformation laws align with the gauge symmetries of Newton-Cartan gravity. 
Both methods have been applied to Einstein gravity and other relativistic generalizations to produce various Newton-Cartan gravity theories \cite{Hartong:2017bwq,Hartong:2022lsy}. 
However, it remains unclear which theory can be considered the ``seed theory" underlying the Chern-Simons TTNC \cite{Hartong:2016yrf} prior to applying these algorithms. 

The novel aspects of our work is to show that the su$(1,2)\oplus$u$(1)$ Chern-Simons theory discussed in this paper serves as the precise candidate for generating the Chern-Simons TTNC upon $(1/c)$-expansion \cite{Hartong:2016yrf}, thereby addressing this long-standing question.
The u$(1)$ extension plays a pivotal role in two key aspects.
First, from the perspective of the four dimensional $\Omega$-background, the u$(1)$ describes the translation along the fibre while su$(1,2)$ is the isometry of $\tilde{\mathbb{C}\mathbb{P}}^2$ transverse to the fibre. 
Therefore, the u$(1)$ arises naturally from the null reduction of so$(2,4)$ \cite{Lambert:2021nol}, suggesting a potential link between our theory and the four-dimensional Lorentzian gravity framework.
Second, as observed in \cite{Chernyavsky:2019hyp}, the extended Schr\"odinger algebra \cite{Hartong:2016yrf} was viewed as the contraction of two copies of su$(1,2)$ algebra, exhibiting a structure analogous to the bosonic super BMS algebra. 
Their work thus constructed a relativistic gravity theory for the extended Schr\"odinger algebra by contracting two su$(1,2)$ Chern-Simons theories (omitting the u$(1)$ extension), and it employs a diagonal embedding of the sl$(2)$ to su$(1,2)$ algebra.
However, the Chern-Simons theory related to the diagonal embedding of su$(1,2)$ algebra was known to be not equivalent to gravitational theories \cite{Lei:2015ika}, as the spin connection cannot be uniquely solved from vielbein.
This issue is resolved by incorporating curvature constraints associated with the u$(1)$ gauge field \cite{Andringa:2010it}.

Beyond these foundational aspects, our work advances the understanding \cite{Chernyavsky:2019hyp} in several novel directions.
The same algebra can support distinct vacua, potentially corresponding to different physical theories.
While \cite{Chernyavsky:2019hyp} reinterprets the Chern-Simons theory by gauging the extended Schr\"odinger algebra in alternative basis as a Poincaré-type gravity, the physical nature of the original Schr\"odinger gravity constructed in \cite{Hartong:2016yrf} and its connection to su$(1,2)$ type of gravity remains unclear despite their shared algebraic structure. 
Therefore, the understanding of Schr\"odinger gravity \cite{Hartong:2016yrf} as an example of $1/c$ expansion of the general frame work proposed in \cite{Hansen:2020pqs}, with the seed theory being su$(1,2)\oplus$u$(1)$ Chern-Simons theory addresses this unresolved issue \footnote{This connection was briefly raised in \cite{Kasikci:2020qsj}. 
However, the omitting of U$(1)$ gauge fields renders the gravitational theory in metric-like fields inequivalent to the gauge-theoretic construction. }. 
Additionally, we discuss how the extended Schr\"odinger algebra \cite{Hartong:2016yrf}—interpreted in \cite{Chernyavsky:2019hyp} as a bosonic analogue of the super Poincaré algebra—can be related to the infinite-symmetry extended algebra derived from the su$(1,2)\oplus$u$(1)$ Chern-Simons theory. 

The remainder of our paper is organized as follows. 
In Section \ref{sec:CSSU12}, we will discuss how the action is computed and a few features of the theory.
We will also use $(1/c)$-expansion perspective to relate our theory to the early works \cite{Hartong:2016yrf}. 
We should view our construction as a novel but also exceptional example of Type I TNC in $2+1$ dimensions which preserves the u$(1)$ particle number \cite{Hansen:2019pkl}.
In Section \ref{sec:infinitesymmetry}, we discuss the infinitely extended symmetry of su$(1,2)\oplus$u$(1)$ in the Chern-Simons theory.
In Section \ref{sec:discussion} we conclude the paper with a few future directions. 

\section{The su$(1,2)\oplus$u$(1)$ Chern-Simons theory }\label{sec:CSSU12}

The Chern-Simons gauge theory is characterized by the following action: 
\begin{equation}
	S[ \mathcal{A}]=\frac{k}{4\pi} \int_M  \text{tr}\left( \mathcal{A}\wedge d \mathcal{A} + \frac{2}{3} \mathcal{A} \wedge  \mathcal{A} \wedge  \mathcal{A}  \right) \,.
\end{equation}
Here, the trace action tr is performed over the specific representation in which the gauge field $\mathcal{A}$ takes its values. 
When the gauge fields are associated with non-semisimple Lie algebras, the formulation of the corresponding Chern-Simons theory is contingent upon the existence of a non-degenerate bilinear product defined in the algebra \cite{Nappi:1993ie}.
Notably, the Bargmann, Newton-Hooke, and Schr\"odinger algebra are examples of non-semisimple Lie algebras. 
To construct the Chern-Simons theory and the corresponding gravitational theories on these algebras, it is necessary to introduce additional generators. 
This extension ensures that the bilinear product on the algebras becomes non-degenerate, thereby enabling a well-defined theory \cite{Papageorgiou:2009zc,Papageorgiou:2010ud,Hartong:2016yrf,Bergshoeff:2016lwr,Hartong:2017bwq}. 

This section will be organized as follows. 
In section \ref{ssec:gaugealgebra}, we will discuss the general structures of the Newton-Cartan gravity emerged from gauging the su$(1,2)\oplus$u$(1)$ algebra.
The discussion is built up at the on-shell level. 
Section \ref{ssec:action} aims to build an action of the torsional Newton-Cartan gravity by the su$(1,2)\oplus$u$(1)$ Chern-Simons theory. 
The notation used for gravitational theories based on Chern-Simons theory is as follows:
\begin{itemize}
	\item In three dimensions, take the ansatz of vielbein as $E_\mu^A$ where $A=0,1,2$. 
	The indices 
	\begin{equation}\label{eq:split-indices}
		A=(0,a)
	\end{equation} 
are split into temporal part $A=0$ and the spatial part $a=1,2$.
	\item We take capital letters as the fields of the Chern-Simons theory. 
	These include $E_\mu^0, E_\mu^a\,,\Omega_\mu^a\,,\Omega_\mu\,,B_\mu$ etc.
	\item We take the usual conventions of Newton-Cartan gravity $\tau_\mu\,, e_\mu^a\,, \omega_\mu^a\,,\omega_\mu$ for the fields after performing $1/c$ expansions
	\item Since the temporal and spatial components of the indices have already been separated, there is no need to alter the sign when raising or lowering the $0$ index. Instead, we adopt the following normalization: 
	\begin{equation}
	-	E_\mu^0 E^\nu_0 + E_\mu^a E_a^\nu =\delta_{\mu}^\nu , \quad \text{and }  \quad E_\mu^0 \equiv E_{\mu,0} \,.
	\end{equation}
\end{itemize}

\subsection{Gauging su$(1,2)\oplus$ u$(1)$}
\label{ssec:gaugealgebra}

We take the following basis of su$(1,2)\oplus$u$(1)$ algebra as it explicitly reveals its connection to the Schr\"odinger algebra, as explained in \cite{Baiguera:2024vlj}. 
We will introduce an extra u$(1)$ generator such that the su$(1,2)\oplus$u$(1)$ algebra is 
\begin{align}\label{eq:newSU12}
	\begin{split}
		& [J,P_a] =\epsilon_{ab} P_b, \qquad [J,G_a] =\epsilon_{ab} G_b, \\
		&[H,K]=D, \quad  \quad [D,K]=2K, \quad [D,H]=-2H \\
		& [K,P_a] =-G_a, \quad [H,G_a] =P_a, \quad [P_a,D]=P_a,\quad [G_a,D]=-G_a \\
		& [P_a,G_b] = \delta_{ab}\left(N- \frac{3}{2}J \right) + \frac{1}{2} \epsilon_{ab} D, \quad [P_a,P_b] =\epsilon_{ab}H,\quad [G_a,G_b] = \epsilon_{ab} K \,.
	\end{split}
\end{align}
Here, $P_a$ with $(a=1,2)$ are spatial translations and $H$ is the time translation. 
$G_a$ are the Galilean boosts and $J$ is the rotation within the spatial directions. 
Except for the dilatation generator $D$, there is an extra special conformal transformation $K$. 
$N$ is the u$(1)$ central generator commutative with the rest of the algebra. 
We observe that the first three lines of \eqref{eq:newSU12} are precisely the $z=2$ Schr\"odinger algebra up to the central element $N$. 
A detailed mapping to the Schr\"odinger algebra is illustrated in \cite{Baiguera:2024vlj}.

The bilinear product of the algebra \eqref{eq:newSU12} is determined by its Cartan-Killing metric denoted as $B(x,y)$.
Thus, the bilinear product is parametrized as:
\begin{align}\label{eq:bilinear-product}
	\begin{split}
		& B(H,K) =B(P_1,G_2) =-B(P_2 ,G_1) =c_{S}, \quad B(D,D) =-2c_{S}\\
		& B(J,J) = \frac{2}{3}(c_{S}+c_U) \quad B(J,N) = c_U \quad B(N,N) = \frac{3}{2}c_U  \,.
	\end{split}
\end{align}
The parameter $c_{S}$ is for the su$(1,2)$ while the $c_U$ is due to the presence of u$(1)$ generator. 
We call the $c_S=-c_U$ limit as the \emph{fully coupled limit}, which will be relevant later for the corresponding gravitational theories. 
This algebra naturally appears as the conformal Killing symmetry of the geometry from the null reduction of the $\Omega$-deformed background \cite{Lambert:2021nol}.
In this paper, we focus on the Chern-Simons theory in the basis \eqref{eq:newSU12}.
There is a manifest $z=2$ Lifshitz subalgebra (spanned by $\{H,D,P_a,J\}$) in this basis, which is relevant to the appearance of $z=2$ Lifshitz geometry in this model, in relation to the Chern-Simons TTNC \cite{Hartong:2016yrf}.

We therefore take the gauge field valued in su$(1,2)\oplus$u$(1)$ algebra as 
\begin{equation}
 \mathcal{A}= H E^0 +P_a E^a + G_a \Omega^a +J \Omega+ D B + K U + N M  \,,
\end{equation}
with the field strength tensors defined as 
\begin{align}
	\begin{split}
\mathcal{F} &= d \mathcal{A} +  \mathcal{A} \wedge  \mathcal{A} \\
& = H R(H) + P_a R^a(P) +G_a R^a(G) + J R(J) + D R(D) +  K R(K) +N R(N)\,,
		\end{split}
\end{align}
which are explicitly 
\begin{equation}\label{eq:curvature-form}
	\begin{aligned}
		&R(H)=d E^0-2B\wedge E^0+ \frac{1}{2}\epsilon_{ab}E^{a}\wedge E^{b} \\
		&R^a(P)=d E^a+E^0\wedge\Omega^a+E^a\wedge B+\epsilon_{ab}E^b\wedge\Omega   \\
		&R^a(G)=d\Omega^a+E^a\wedge U-\Omega^a\wedge B+\epsilon_{ab}\Omega^b\wedge\Omega  \\
		&R(J)=d\Omega-\frac{3}{2}E^{a}\wedge\Omega^{a}  \\
		&R(D)=d B+E^0\wedge U+\frac{1}{2}\epsilon_{ab}E^a\wedge\Omega^b  \\
		&R(K)=dU+2B\wedge U + \frac{1}{2}\epsilon_{ab}\Omega^{a}\wedge\Omega^{b}\\
		&R(N)=dM+E^a\wedge \Omega^a  \,.
	\end{aligned}
\end{equation}
The inverse vielbein are defined as 
\begin{equation}\label{eq:def-inverse-vielbein}
E_\mu^0 E^\mu_0 = -1 , \quad  E_\mu^a E^\mu_b =\delta_b^a, \quad E_\mu^a E^\mu_0 =0, \quad E_\mu^0 E^\mu_a =0 \,.
\end{equation}
The gauge theory is related to gravitational theory by imposing curvature constraints. 
These constraints include \cite{Hartong:2016yrf,Bergshoeff:2014uea}
\begin{equation}\label{eq:curvature-constraint}
	R(H)=R^a(P) = R(N)=R(D)=0 \,.
\end{equation}
Additional constraints can be found by considering the Bianchi identities. 
In particular, similar to the models obtained by gauging the Bargmann algebra and the Schr\"odinger algebra, the constraints $R(H)=R^a(P)=R(N)=0$ can be used to uniquely determine the spin-connection from the vielbein $E_\mu^0,E_\mu^a$ and the gauge field $M_\mu$ together with the $E^\mu_0 B_\mu$. 
\begin{equation}
	\begin{aligned}
		\Omega_{\mu}&=\epsilon_{ab}(E^{\nu a}\partial_{[\mu}E_{\nu]}{}^{b}-E^{\nu b}\partial_{[\mu}E_{\nu]}{}^{a}+E_{\mu}{}^{c}\partial_{[\nu}E_{\rho]}{}^{c}E^{\nu a}E^{\rho b}\\
		&-E^0_{\mu}E^{\rho b}E^{\nu a}\partial_{[\nu}M_{\rho]}+2E_\mu^b E^{\nu a}E^\rho_0\partial_{[\nu}E_{\rho]}^0), \\
		\Omega_{\mu}{}^{a}&=E^{\nu a}\partial_{[\mu}M_{\nu]}+E^{\rho a}E_0^{\nu}E_{\mu b}\partial_{[\nu}E_{\rho]}{}^{b}+E^0_{\mu}E_0^{\rho}E^{\nu a}\partial_{[\nu}M_{\rho]}\\
		&+E_0^{\nu}\partial_{[\mu}E_{\nu]}{}^{a}-(E^\nu_0B_\nu)E_\mu^a\,.\\
	\end{aligned}
\end{equation}
The constraints \eqref{eq:curvature-constraint} exactly resolve the problem that the su$(1,2)$ Chern-Simons theory is degenerate \cite{Lei:2015ika}; namely, the spin connection is not determined uniquely from vielbein \footnote{The work \cite{Kasikci:2020qsj} did not introduce this u$(1)$ generator, thus the corresponding gravity theory is still not equivalent to the Chern-Simons gauge theory.}.
A notable distinction between the models discussed in the literature  \cite{Hartong:2016yrf, Bergshoeff:2014uea,Hartong:2015zia,Andringa:2010it} and the models described in \eqref{eq:curvature-form} lies in the curvature $R(H)=0$.
This condition signifies that the theory corresponds to the torsional Newton-Cartan gravity as $E^0 \wedge dE^0 = E^0 \wedge E^1 \wedge E^2 \neq 0$.
The presence of the special torsion structure is essential for enhanced conformal symmetry su$(1,2)$. 
We will also briefly discuss the five dimensional example in Appendix \ref{appendix:SU13}, which has the symmetry su$(1,3)\oplus$u$(1)$. 

The gauge field $\mathcal{A}$ transform under the the internal symmetries $G_a,J,D,K,N$ as follows 
\begin{equation}
	\delta \mathcal{A}_\mu=\partial_\mu\Sigma+[\mathcal{A}_\mu,\Sigma], \qquad 
\Sigma=G_a\Lambda^a+J\Lambda+D\Lambda_D+K\Lambda_K+N\sigma \,.
\end{equation}
After incorporating the diffeomorphism transformation, we introduce a new transformation as \cite{Hartong:2015zia}:
\begin{equation}
	\bar{\delta} \mathcal{A}_\mu=\mathcal{L}_{\xi}\mathcal{A}_{\mu}+\partial_\mu\Sigma+[\mathcal{A}_\mu,\Sigma]
\end{equation}
In terms of each component, the gauge transformation of the fields are 
\begin{equation}\label{eq:SU12-Variation}
	\begin{aligned}
		&	\bar{\delta} E^0_{\mu}=\mL_\xi E_\mu^0+ 2E^0_{\mu}\Lambda_{D}  \\
		& 	\bar{\delta} E_{\mu}{}^{a}=\mL_\xi E_\mu^a+E^0_\mu\Lambda^{a}+E_\mu^a\Lambda_D+\epsilon_{ab}E_\mu^b\Lambda  \\
		&	\bar{\delta} \Omega_{\mu}{}^{a}=\mL_\xi \Omega_\mu^a+
		\partial_{\mu}\Lambda^{a}+E_{\mu}^{a}\Lambda_{K}-\Omega_{\mu}^{a}\Lambda_{D}+B_{\mu}\Lambda^{a}+\epsilon_{ab}\Omega_{\mu}^{b}\Lambda-\epsilon_{ab}\Omega_{\mu}\Lambda^{b}  \\
		&	\bar{\delta} \Omega_{\mu} =\mL_\xi \Omega_\mu+\partial_{\mu}\Lambda-\frac{3}{2}E_{\mu}^{a}\Lambda^{a} \\
		&	\bar{\delta} B_{\mu} =\mL_\xi B_\mu+ \partial_{\mu}\Lambda_{D}+E^0_{\mu}\Lambda_{K}+\frac{1}{2}\epsilon_{ab}E_{\mu}^{a}\Lambda^{b}  \\
		&	\bar{\delta} U_{\mu} =\mL_\xi U_\mu+\partial_{\mu}\Lambda_{K}+\epsilon_{ab}\Omega_{\mu}^{a}\Lambda^{b}-2U_{\mu}\Lambda_{D}+2B_{\mu}\Lambda_{K}\\
		&	\bar{\delta} M_\mu=\mL_\xi M_\mu +\partial_\mu\sigma+E_{\mu}^a\Lambda^a. 
	\end{aligned}
\end{equation}
The transformations of the inverse vielbein are fixed by combining \eqref{eq:SU12-Variation} with \eqref{eq:def-inverse-vielbein}
\begin{equation}
	\bar{\delta} E_0^\mu=\mL_\xi E^\mu_0-2\Lambda_{D}E_0^\mu+\Lambda^{a}E^{\mu}_{a},\qquad 	\bar{\delta} E^{\mu}_{a}=\mL_\xi E^\mu_a-\Lambda_{D}E^{\mu}_{a}+\epsilon_{ab}\Lambda E^{\mu b}.
\end{equation}

The transformation rules reveal several intriguing features when compared to the twistless torsional Newton-Cartan gravity derived from gauging the $z=2$ Schr\"odinger algebra \cite{Hartong:2016yrf, Bergshoeff:2014uea}.  
First, the curvature constraints $R(H)=R^a(P)=0$ establish a connection between diffeomorphism and gauge transformations.
This feature aligns with the behavior observed in twistless torsional Newton-Cartan gravity theories.
For instance, 
\begin{align}\label{eq:gauge-to-diffeo}
	\begin{split}
\delta_\xi E_\mu^0 &= \xi^\nu \partial_\nu E_\mu^0 + \partial_\mu \xi^\nu  E_\nu^0 \\
&= \partial_\mu(\xi^\nu E_\nu^0) + 2(\xi^\nu B_\nu) E_\mu^0 -2 (\xi^\nu E_\nu^0) B_\mu - \epsilon_{ab} (E_\nu^a \xi^\nu) E_\mu^b -\xi^\nu R_{\mu\nu}(H) \,.
	\end{split}
\end{align}
The curvature constraint $R(H)=0$ then identifies $\xi^\nu B_\nu$ as the dilatation transformation acting on $E_\mu^0$, while  the $\xi^\nu E_\nu^0$ and $\xi^\nu E_\nu^a$ correspond to temporal and spatial translations, respectively.
Despite this similarity,  a notable difference is worth highlighting. 
Both the $B_\mu$ and $\Omega_\mu$ fields transform non-trivially under the Galilean boost $G_a$.
This behavior originates from the $[P_i,G_j]$ commutator in the algebra \eqref{eq:newSU12}.
In contrast to the Schr\"odinger algebra, where $[P_i,G_j]=\delta_{ij} N$ (with no dependence on $D,J$ generators on the RHS), 
the $[P_i,G_j]$ commutators in \eqref{eq:newSU12} exhibit a non-trivial dependence on the $D,J$ generators.
This modification directly leads to the non-trivial behavior of $B_\mu$ and $\Omega_\mu$ under the Galilean boost transformation, as mentioned earlier.
As we will show in the rest of this section, the affine connection exhibits novel transformation structures, distinguishing it from those observed in TTNC/TNC theories derived from gauging the Schr\"odinger algebra \cite{Bergshoeff:2014uea,Hartong:2015zia}. 

\subsection*{Affine connection}

The affine connection is introduced to render derivatives covariant under diffeomorphism transformations.
For completeness, we briefly recall the relevant discussion of affine connections in torsionless Newton–Cartan (NC) and twistless torsional Newton–Cartan (TTNC) theories; a more detailed review is provided in Appendix~\ref{sec:review-NC-gauge}.
Further technical background can be found in references \cite{Bergshoeff:2014uea,Hartong:2016yrf}.
In what follows, we simply collect the key formulae that will be needed in our later discussion.

To begin, we recall from Appendix~\ref{sec:review-NC-gauge} several relations characterizing the affine connections in torsionless NC and TTNC geometries.
In the torsionless NC case, the covariant derivatives are defined as
\begin{align}\label{eq:covariant-Torless-NC-text}
	\begin{split}
		\mathcal{D}_\mu \tau_\nu & = \partial_\mu \tau_\nu -\Gamma_{\mu \nu}^{\rho} \tau_\rho \,, \\
		\mathcal{D}_\mu e_\nu^a & =\partial_\mu e_\nu^a   -\Gamma_{\mu \nu}^{\rho} e_\rho^a -\omega_\mu^a \tau_\nu -\omega_\mu \epsilon_{ab} e_\nu^b  \,.
	\end{split}
\end{align}
The vielbein postulates $\mD_\mu \tau_\nu = \mD_\mu e_\nu^a = 0$ determine the affine connection in the form
\begin{equation}\label{eq:affine-torsionless-frame-text}
	\Gamma_{\mu\nu}^\rho = -v^\rho \partial_\mu\tau_\nu + e^\rho_a (\partial_\mu e_\nu^a -\omega_\mu^a \tau_\nu -\omega_\mu \epsilon^{ab} e_\nu^b)\,.
\end{equation}
Alternatively, one can express the affine connection in terms of the metric-like invariant fields $\tau_\mu$ and $\bar{h}_{\mu\nu}$:
\begin{equation}\label{eq:affine-torsionless-metric-text}
	\Gamma_{\mu\nu}^\rho = -\hat{v}^\rho \partial_\mu \tau_\nu + \frac{1}{2} h^{\rho \sigma} (\partial_\mu \bar{h}_{\sigma \nu} + \partial_\nu \bar{h}_{\mu\sigma} -\partial_\sigma \bar{h}_{\mu\nu}) \,,
\end{equation}
where inverse metric like fields $h^{\mu\nu}$ and $\hat{v}^\mu$ are contrained to satisfy the completeness and orthogonality conditions.
Both expressions \eqref{eq:affine-torsionless-frame-text} and \eqref{eq:affine-torsionless-metric-text} are invariant under all internal gauge transformations of the Bargmann algebra \eqref{eq:gauge-bargmmann}.
Since the formulation of Newton–Cartan gravity in terms of Galilean-invariant metric-like fields is independent of the specific choice of affine connection, the corresponding Riemann tensor obtained from \eqref{eq:affine-torsionless-frame-text} takes the form~\cite{Hartong:2015zia}
\begin{equation}
	R_{\mu\nu\sigma}{}^{\rho} = e^\rho_a \tau_\sigma R_{\mu\nu}^a(G) - e_{\sigma}^a e^\rho_b R_{\mu\nu}(J) \epsilon_{ab}\,,
\end{equation}
where $R(G)$ and $R(J)$ denote the $G$- and $J$-components of the curvature two-form $F= d\mathcal{A} + \mA \wedge \mA$. 

Similarly, in TTNC geometry obtained by gauging the $z=2$ Schr\"odinger algebra, one may introduce dilatation-covariant generalized derivatives, 
\begin{align}\label{eq:covariant-TTNC-text}
	\begin{split}
		\mathcal{D}_\mu \tau_\nu & = \partial_\mu \tau_\nu -\tilde{\Gamma}_{\mu \nu}^{\rho} \tau_\rho-2 b_\mu \tau_\nu \,, \\
		\mathcal{D}_\mu e_\nu^a & =\partial_\mu e_\nu^a   -\tilde{\Gamma}_{\mu \nu}^{\rho} e_\rho^a -\omega_\mu^a \tau_\nu -\omega_\mu \epsilon_{ab} e_\nu^b - b_\mu e_\nu^a  \,,
	\end{split}
\end{align}
which are solved by
\begin{align}\label{eq:affine-TTNC-frame-text}
	\begin{split}
		\tilde{\Gamma}_{\mu\nu}^\rho = -\hat{v}^\rho (\partial_\mu \tau_\nu-2 b_\mu \tau_\nu) + e^\rho_a (\partial_\mu e_\nu^a -\omega_\mu^a \tau_\nu -\omega_\mu \epsilon_{ab} e_\nu^b -b_\mu e_\nu^a)\,.
	\end{split}
\end{align}
Expressed in terms of metric-like fields, the dilatation-covariant affine connection takes the form
\begin{align}\label{eq:affine-TTNC-metric-text}
	\begin{split}
		\tilde{\Gamma}_{\mu\nu}^\rho = -\hat{v}^\rho (\partial_\mu \tau_\nu-2 b_\mu \tau_\nu) + \frac{1}{2} h^{\rho \sigma} \Big[(\partial_\mu-2b_\mu) \bar{h}_{\nu\sigma} +  (\partial_\nu-2b_\nu) \bar{h}_{\mu\sigma}- (\partial_\sigma-2b_\sigma) \bar{h}_{\mu\nu}\Big]
	\end{split}
\end{align}
which is invariant under all $G_a,J,N,D$ transformations. 
Both \eqref{eq:affine-TTNC-frame-text} and \eqref{eq:affine-TTNC-metric-text} transform nontrivially under the special conformal transformation $K$  \footnote{This indicates the difference between \eqref{eq:affine-TTNC-metric-text} and \eqref{eq:spec-trans-affine-TTNC-text} is a tensor invariant under all the internal gauge transformations. 
To find the difference, we need to solve the curvature constraints explicitly following \cite{Bergshoeff:2014uea,Hartong:2015zia,Hartong:2016yrf}. 
However, we can use any of these choices of affine connection to construct curvature invariants.}
\begin{equation}\label{eq:spec-trans-affine-TTNC-text}
	\bar{\delta} \tilde{\Gamma}_{\mu\nu}^\rho = -\Lambda_K (\delta_\mu^\rho \tau_\nu + \delta_\nu^\rho \tau_\mu)\,. 
\end{equation}
This nontrivial $K$-transformation implies that the associated Riemann tensor inevitably contains a nontrivial manifest dependence on the gauge field $f_\mu$ introduced by gauging $K$ \cite{Hartong:2016yrf,Hartong:2015zia}: 
\begin{equation}\label{eq:Riemmantensor-TTNC}
	\tilde{R}_{\mu\nu\sigma}{}^{\rho}= -e^{\rho d} e_\sigma^c \tilde{R}_{\mu\nu}(J) +e^\rho_c \tau_\sigma \tilde{R}_{\mu\nu}^c (G) - \delta_\mu^\rho  f_\nu \tau_\sigma + \delta_\nu^\rho  f_\mu \tau_\sigma + \delta_\sigma^\rho (f_\mu\tau_\nu-f_\nu \tau_\mu)
\end{equation}
Thus, in the TTNC framework, the $f_\mu$ dependence of the Riemann tensor directly reflects the fact that the affine connection is not invariant under $K$.
 
On the other hand, in the TTNC theory one may also employ the ordinary affine connection $\Gamma_{\mu\nu}^\rho$ from \eqref{eq:affine-torsionless-metric-text} to define the covariant derivative.
The difference between $\Gamma_{\mu\nu}^\rho$ and $\tilde{\Gamma}_{\mu\nu}^\rho$ is a tensor under both internal gauge transformations and diffeomorphisms, and either choice can be used consistently.
Each, however, has its own advantages.
For example, the dilatation-generalized connection $\tilde{\Gamma}_{\mu\nu}^\rho$ is not metric compatible, i.e. it does not satisfy
\begin{equation}
	\nabla_\mu \tau_\nu = \nabla_\mu h^{\mu\nu} =0
\end{equation}
whereas the ordinary connection $\Gamma_{\mu\nu}^\rho$ does. 
On the other hand, $\Gamma_{\mu\nu}^\rho$ lacks dilatation gauge invariance.
Consequently, when using $\Gamma_{\mu\nu}^\rho$ in place of $\tilde{\Gamma}_{\mu\nu}^\rho$, an explicit $b\mu$ dependence must be included in the Riemann tensor \eqref{eq:Riemmantensor-TTNC}. 
In contrast, the connection $\tilde{\Gamma}_{\mu\nu}^\rho$ restores dilatation invariance, which simplifies the expression for the Riemann tensor \eqref{eq:Riemmantensor-TTNC} by removing its manifest dependence on $b_\mu$.
Indeed, one can enhance the dilatation invariance of $\Gamma_{\mu\nu}^\rho$ itself by replacing each ordinary partial derivative with a dilatation-covariant derivative.

Finally, in the TNC case we begin with the affine connection constructed from the vielbein.
As in the TTNC discussion, one may either use the dilatation-covariant generalized connection $\tilde{\Gamma}_{\mu\nu}^\rho$ in \eqref{eq:affine-TNC-covar}, or employ the ordinary connection $\Gamma_{\mu\nu}^\rho$ obtained by replacing the dilatation-covariant derivatives in \eqref{eq:affine-TNC-covar} with ordinary partial derivatives. 
The dlatation covariant derivatives are then defined by  $\tilde{\Gamma}_{\mu\nu}^{\rho}$ as
\begin{align}\label{eq:affine-TNC-covar}
	\begin{split}
		\tilde{\mathcal{D}}_\mu E_\nu^0 &\equiv \partial_\mu E_\nu^0 - \tilde{\Gamma}_{\mu\nu}^\rho E_\rho^0 -2B_\mu E_\nu^0\,,\\
		\tilde{\mathcal{D}}_{\mu}{E_{\nu}}^{a}& \equiv\partial_{\mu}E_{\nu}^{a}-\tilde{\Gamma}_{\mu\nu}^{\rho}E_{\rho}^{a} -B_\mu E_\nu^a-\Omega_{\mu}^{a}E^0_{\nu}-\epsilon_{ab}\Omega_{\mu}  E_{\nu}^{b}.
	\end{split}
\end{align}
The vielbein postulates determine the affine connection as 
\begin{equation}\label{eq:affine-with-B-vielbein}
		\tilde{\Gamma}_{\mu\nu}^{\rho}=-E_0^{\rho}(\partial_{\mu}-2B_{\mu})E^0_{\nu}+E^{\rho}_a\Big[(\partial_{\mu}-B_\mu)E_{\nu}{}^{a}-\Omega_{\mu}{}^{a}E^0_{\nu}-\epsilon_{ab}\Omega_{\mu}{}E_{\nu}^b \Big] \,.
\end{equation} 
Although \eqref{eq:affine-with-B-vielbein} is formally identical to the expression obtained by gauging the Schr\"odinger algebra, its transformation under internal gauge symmetries is different in the Galilean boost transformation:
\begin{equation} \label{eq:affine-with-Bvari-gauge}
\delta\tilde{\Gamma}^\rho_{\mu\nu}=\Lambda^a \left(\epsilon_{ab}E_\mu^b \delta^\rho_\nu-\frac{1}{2}\epsilon_{ab}E_\mu^b E_\nu^c E^\rho_c-\frac{3}{2}\epsilon_{bc}E_\mu^a E_\nu^b E^{\rho c} \right)
    -\Lambda_K(E_\mu^0 \delta^\rho_\nu+E_\nu^0 \delta^\rho_\mu) \,.
\end{equation} 
The loss of boost invariance in \eqref{eq:affine-with-Bvari-gauge} originates from the fields $B_\mu,\Omega_\mu$ fields, which transform nontrivially under Galilean boosts.
This distinguishes our setup from TTNC gravity obtained by gauging the Schr\"odinger algebra \cite{Hartong:2015zia} and from the torsional Newton–Cartan framework \cite{Bergshoeff:2014uea}.
The Riemann tensor evaluates to
\begin{equation}\label{eq:Riemman-su12}
	\begin{aligned}
		\tilde{R}_{\mu\nu\sigma}{}^{\rho}=&E_{c}^{\rho}E^0_{\sigma}R_{\mu\nu}^{c}(G)  -\epsilon_{ab}E^{\rho a}E_{\sigma}^{b}R_{\mu\nu}(J) -\delta_{\mu}^{\rho}E^0_{\sigma}U_{\nu}+\delta_{\nu}^{\rho}E_{\sigma}^0U_{\mu}+\delta_{\sigma}^{\rho}\left(U_{\mu}E^0_{\nu}-U_{\nu}E^0_{\mu}\right)\\
		&+\epsilon_{ab}\delta_\sigma^\rho E_{[\mu}^a\Omega_{\nu]}^b+3\epsilon_{ab}E^\rho_0 E_\sigma^0 E_{[\mu}^a\Omega_{\nu]}^b+3\epsilon_{ab}E^{\rho a}E_\sigma^b E_{[\mu}^c\Omega_{\nu]}^c \,.
	\end{aligned}
\end{equation}
This behavior is analogous to \eqref{eq:Riemmantensor-TTNC}.
As expected, the explicit appearance of $U_\mu,\Omega_\mu^a$ fields shows that the affine connection is not invariant under either Galilean boosts or the special conformal transformation i.e. \eqref{eq:affine-with-Bvari-gauge}.
Moreover, because $U_\mu,\Omega_\mu^a$ themselves transform nontrivially under boosts, the ordinary connection $\Gamma_{\mu\nu}^\rho$ 
is not boost-invariant either. 
Within the above vielbein-based ansatz (local and first-order in derivatives), we have not found a choice of affine connection that remains invariant under Galilean boosts. 

On the other hand, we can define boost-invariant metric-like fields.
In analogy with torsionless NC and TTNC, both $E_\mu^0$ and $h^{\mu\nu} = \delta_{ab} E^\mu_a E^\nu_b$ are boost invariant. 
are boost invariant.
Using the U$(1)$ gauge field $M_\mu$, we further define the boost invariants
\begin{equation}\label{eq:def-hatE-barh}
\hat{E}^\mu_0 =E^\mu_0 - h^{\mu\nu} M_\nu, \qquad \bar{h}_{\mu\nu}=h_{\mu\nu} -E_\mu^0 M_\nu -E_\nu^0 M_\mu  \,.
\end{equation}
Accordingly, the following affine connection is manifestly boost invariant:
\begin{equation}\label{eq:boostinvariant-affine}
	\bar{	\Gamma}_{\mu\nu}^\rho = -\hat{E}_0^\rho \partial_\mu E_\nu^0 + \frac{1}{2} h^{\rho \alpha} \left(\partial_\mu \bar{h}_{\nu\alpha} +\partial_\nu \bar{h}_{\mu\alpha} - \partial_\alpha \bar{h}_{\mu\nu}  \right)  \,.
\end{equation}
However, this connection is not invariant under dilatations.
As in the TTNC case, replacing 
$\partial_\mu$ by $\partial_\mu-2B_\mu$ restores dilatation invariance, but the appearance of $B_\mu$ —which transforms nontrivially under Galilean boosts—then breaks boost invariance.
Therefore, unlike in TTNC, a metric-like affine connection cannot simultaneously preserve both boost and dilatation invariance in this construction.
For example, \eqref{eq:boostinvariant-affine} transforms as
\begin{align}
	\begin{split}
\delta\bar{\Gamma}_{\mu\nu}^\rho&= -2\hat{E}^\rho_0 E_\nu^0\partial_\mu\Lambda_D+h^{\rho\alpha}(\bar{h}_{\mu\alpha}\partial_\nu\Lambda_D+\bar{h}_{\nu\alpha}\partial_\mu\Lambda_D-\bar{h}_{\mu\nu}\partial_\alpha\Lambda_D)\\
&+\frac{1}{2}h^{\rho\alpha}[(\partial_\mu E_\nu^0-\partial_\nu E_\mu^0)\partial_\alpha\sigma+(\partial_\alpha E_\mu^0-\partial_\mu E_\alpha^0)\partial_\nu\sigma+(\partial_\alpha E_\nu^0-\partial_\nu E_\alpha^0)\partial_\mu\sigma]
	\end{split}
\end{align} 
This conflict does not arise in TTNC, since the dilatation gauge field $b_\mu$ is invariant under Galilean boosts.
The difference ultimately traces back to the nontrivial appearance of the $D,J$ generators in the $[P_i,G_j]$ commutator. 
In principle, the physical theory is independent any of the above choices of affine connections, although different choices lead to different formulations.

\subsection{Torsional Newton-Cartan gravity}
\label{ssec:action}

In higher dimensions, the construction of Newton-Cartan gravity actions typically includes methods such as the null reduction and $1/c$ expansion of relativistic gravitational theories \cite{Hansen:2020pqs}. 
In three dimensions, the Chern-Simons theory serves as a powerful tool for constructing gravitational actions, as demonstrated in \cite{Witten:1988hc}. 
The su$(1,2)\oplus$u$(1)$ Chern-Simons action with the bilinear product \eqref{eq:bilinear-product} can be written as 
\begin{eqnarray}\label{eq:CS-SU12}
&& \text{tr} \left( \mathcal{A}\wedge d \mathcal{A} + \frac{2}{3} \mathcal{A} \wedge  \mathcal{A} \wedge  \mathcal{A}  \right) \\ \nonumber
&= & 2c_S\Big[\Omega^2 \wedge R^1(P) -\Omega^1 \wedge R^2(P) +U\wedge R(H)  -E^0 \wedge \Omega^1 \wedge \Omega^2 -B\wedge dB +\frac{1}{3}\Omega \wedge d\Omega \Big] \\ \nonumber
&&+c_U \left( 2M\wedge d\Omega+\frac{3}{2}M\wedge d M +\frac{2}{3}\Omega\wedge d\Omega\right)  \,.
\end{eqnarray}
where the curvature forms are computed in \eqref{eq:curvature-form}. 
The spin connection $\Omega^a$ works as the Lagrange multiplers to impose the curvature constraint $R^a(P)=0$ while the field $U$ related to special conformal transformation work as the Lagrange multipler to impose the constraint $R(H)=0$.
In addition, the U$(1)$ particle number invariance is automatically preserved in \eqref{eq:CS-SU12}, which can be seen as another type I TNC that preserves the u$(1)$ symmetry, in addition to the examples found \cite{Hartong:2016yrf,Bergshoeff:2016lwr}.

The action \eqref{eq:CS-SU12} can be related to the extended Schr\"odinger gravity built up in \cite{Hartong:2016yrf} by $1/c$ expansion \cite{Kasikci:2020qsj}.
Namely, taking the following ansatz for non-relativistic expansion,
\begin{equation}
	\begin{aligned}
		&E^{0} =c\tau+\beta, \quad U=\frac{1}{c}f+\frac{1}{c^{2}}\zeta, \quad M=\frac{1}{c}m+\mathcal{O}(c^{-2})  \\
		&B=b-\frac{1}{2c}\alpha,\quad\Omega=\omega+\mathcal{O}(c^{-1})  \\
		&E^{i} =e^i+\mathcal{O}(c^{-1}),\quad \Omega^{i}=\frac{1}{c}\omega^{i}+\mathcal{O}(c^{-2})\,,
	\end{aligned}
\end{equation} 
we can exactly acquire the action  \cite{Hartong:2016yrf}
\begin{equation}\label{eq:Schrodinger-gravity}
	\begin{aligned}
		\mathcal{L}&=2c_1\Big[e^1\wedge R^2(G)-e^2\wedge R^1(G)+\tau\wedge\omega^1\wedge\omega^2-f\wedge e^1\wedge e^2\\
		&+\zeta\wedge(d\tau-2b\wedge\tau)+\alpha\wedge(db-f\wedge\tau)+\beta\wedge(df+2b\wedge f)-m\wedge d\omega \Big]\\
		&+2c_2[b\wedge db-\tau\wedge df-2b\wedge f\wedge\tau] +c_3\omega\wedge d\omega\,.
	\end{aligned}
\end{equation}
by identifying
\begin{equation}
	c_1 = \frac{c_S}{c}, \quad c_2= c_S, \quad c_3 = \frac{2c_S}{3} +2c_U \,.
\end{equation}
Parameters $c_1,c_2,c_3$ are three non-vanishing coefficients permitted in the non-degenerate bilinear product through the gauging of the extended Schr\"odinger algebra \cite{Hartong:2016yrf}. 
In the corresponding Schr\"odinger gravity \eqref{eq:Schrodinger-gravity}, the terms associated with $c_1$ give rise to the dynamical TTNC action, supplemented by constraints enforced by the Lagrange multiplers $\zeta,\alpha,\beta$ and the U$(1)$ gauge field $m$. 
Meanwhile, the equations of motion derived from the terms associated to $c_2,c_3$ reproduce the constraints of the TTNC action. 
Notably, the $c_2,c_3$ appear in the leading order in the $1/c$ expansion while the main part of the dynamical gravity is encoded with the parameter $c_1$ as the subleading order. 
These two features closely parallel the $1/c$ expansion of general relativity in the derivation of Newton-Cartan gravity \cite{Hansen:2020pqs,Hansen:2020wqw}, where the subleading order equation of motion always contains the equation of motion of the previous orders. 
The generators of the extended Schr\"odinger algebra can be further acquired by \cite{Baiguera:2024vlj}
\begin{equation}\label{eq:expansion-Aguage}
	\mathcal{A}=	\mathcal{A}^I T_I = \sum_{n=0}^\infty \overset{(n)}{\mathcal{A}^I} c^{-n} T_I\,, \quad  T_I^{(n)} = T_I \otimes c^{-n} \, ,
	\quad%
	T_I=\{H,D,K,J\}\,.
\end{equation}

We will then discuss a few features of \eqref{eq:CS-SU12}.
The action can be formulated by metric like fields after imposing the curvature constraints \eqref{eq:curvature-constraint}. 
The part proportional to $c_S$ is reminiscent of the conformal gravity in three dimensions \cite{Deser:1981wh,Horne:1988jf,Merbis:2014vja,Afshar:2011yh,Afshar:2011qw,Afshar:2014rwa,Witten:2007kt,Fuentealba:2020zkf,Fuentealba:2024thk}, which can be acquired by gauging so$(2,3)$ algebra.
The formulation is briefly reviewed in Appendix \ref{appendix:SO32}.
This algebra includes two additional special conformal transformation generators $K^a$ and we will denote the corresponding gauge fields as $F^a$. 
The theory expressed in terms of the vielbein formalism is:
\begin{align} \label{eq:SO32-veilbein-action}
	\begin{split}
		\frac{\mathcal{L}_{\text{so}(3,2)}}{c_S} &= -2F \wedge R(H) -2 F^a \wedge R^a (P) \\
		& -\Omega \wedge d\Omega + \Omega^a \wedge d \Omega^a +2\Omega \wedge \Omega^1 \wedge \Omega^2+ B \wedge dB  \,.
	\end{split}
\end{align} 
Remarkably, action \eqref{eq:SO32-veilbein-action} is equivalent to the gravitational Chern-Simons action modulo boundary terms \cite{Kraus:2005zm}. 
The explicit difference is given by 
\begin{equation}\label{eq:difference-CS-gCS}
	S_{\text{so}(3,2)}[A] - S_{gCS} = \frac{k}{12\pi} \int_M \text{tr}(E^{-1}dE)^3 - \frac{k}{4\pi} \int_{\partial M} \text{tr}(\Omega dE E^{-1})  \,,
\end{equation}
where the gravitational Chern-Simons term $S_{gCS}$ takes the standard form:
\begin{equation}
	S_{gCS} = \int d^3 x \epsilon^{\mu\nu\rho}\Gamma_{\mu\sigma}^{\lambda}\left(\partial_{\nu}\Gamma_{\rho\lambda}^{\sigma}+\frac{2}{3}\Gamma_{\nu\tau}^{\sigma}\Gamma_{\rho\lambda}^{\tau}\right)\,. 
\end{equation}

To observe the similarity between su$(1,2)\oplus$u$(1)$ Chern-Simons theory and the Lorentzian conformal gravity \eqref{eq:SO32-veilbein-action}, we would focus on the fully coupled limit at $c_S=-c_U$ of \eqref{eq:CS-SU12}, where the action can be simplified after imposing the curvature constraints (up to boundary terms): 
\begin{align}\label{eq:CS-SU12-2}
	\begin{split}
	& \text{tr} \left( \mathcal{A}\wedge d \mathcal{A} + \frac{2}{3} \mathcal{A} \wedge  \mathcal{A} \wedge  \mathcal{A}  \right) \\ 
	=&  2c_S\Big[ -E^0 \wedge \Omega^1 \wedge \Omega^2   +E^a \wedge \Omega^a \wedge \Omega+\frac{3}{4}M\wedge d M-B\wedge dB \Big] \,.
	\end{split}
\end{align}
The action \eqref{eq:CS-SU12-2} consists of three parts. 
The first two terms, which also appear in Schr\"odinger gravity \cite{Hartong:2016yrf}, can be reformulated in terms of metric formulation, by integrating out the connections $\Omega^a,\Omega,B,U$. 
The $M \wedge dM$ can be interpreted as the Chern-Simons theory for a U$(1)$ gauge field. 
See \cite{Hartong:2017bwq} for an example with U$(1)$ gauge field and relations to non-relativistic gravitational theory. 
It is worth noting that after imposing the curvature constraints, the only dynamic component of the $B_\mu$ field is $E^\mu_0 B_\mu$.
The last term, $B \wedge dB$ can be gauged away by choosing the appropriate special conformal transformations $\Lambda_K$.
For a detailed derivation of the metric-like descriptions of the action \eqref{eq:CS-SU12-2} we refer to \cite{Hartong:2016yrf}  and will not reiterate it here.

\subsection*{Vacuum solution}

Before we end this section, we would like to discuss the vacuum solution to the su$(1,2)\oplus$u$(1)$ Chern-Simons theory.
Owing to its topological nature, all the solutions to the flatness condition 
\begin{equation} \label{eq:equationofmotion}
	F= d\mathcal{A}+ \mathcal{A}\wedge \mathcal{A}=0
\end{equation}
are locally equivalent and related by gauge transformation.
However, there are interesting solutions in specific gauges. 
To display this example, let the spacetime coordinates be denoted as $t,x_{a}$, where $a=1,2$. 
In polar coordinates, these are expressed as
\begin{equation}\label{eq:rtheta-x1x2}
	x_1= r \cos\theta, \qquad x_2 =r \sin \theta, \qquad \theta\in [0,2\pi]  \,.
\end{equation}
An interesting solution can be formulated in polar coordinates as
\begin{equation}\label{eq:vacuumsolution-SU12}
\mathcal{A}_{\text{vac}} = H \left(\frac{dt}{r^2} -\frac{1}{2}d\theta\right) + P_1 \frac{d x_1}{r} + P_2\frac{dx_2}{r} - \frac{dr}{r} D\,,
\end{equation}
Of course \eqref{eq:vacuumsolution-SU12} is not a fully simplified form as $x_1,x_2$ are related to $r,\theta$ by \eqref{eq:rtheta-x1x2}. 
However, this form is convenient for two reasons. 

First, it is known that Newton-Cartan geometry can be acquired from the null reduction of a Lorentzian manifold. 
Namely, given $E_\mu^0,E_\mu^a$, the gauge field $M_\mu$ and the null coordinate $u$ 
\begin{equation}\label{eq:null-redu-ansatz}
	ds^2 = -2E_\mu^0(du-M_\mu dx^\mu) +h_{\mu\nu} dx^\mu dx^\nu \,.
\end{equation}
The geometry generating the solution \eqref{eq:vacuumsolution-SU12} is described by the metric
\begin{equation}\label{eq:def-ds4}
	ds^2 = \frac{1}{r^2} ds_{4}^2 = \frac{1}{r^2} \left[
2du \left(dt-\frac{r^2}{2}d\theta \right) + dr^2+ r^2 d\theta^2	
	\right]\,,
\end{equation}
which is conformal to a four-dimensional manifold defined by $ds_4^2$. 
This geometry corresponds to a torsional Newton-Cartan geometry with su$(1,2)$ symmetry in the operator picture \cite{Baiguera:2024vlj,Lambert:2021nol}.
This suggests that the su$(1,2)\oplus$u$(1)$ Chern-Simons theory under investigation may have a potential connection to a four-dimensional gravitational theory, whose vacuum solution can be taken as $ds_4^2$. 

Second, as mentioned in \cite{Baiguera:2024vlj}, the algebra \eqref{eq:newSU12} reduces to the Schr\"odinger algebra by rescaling 
\begin{equation}
	P_a \to c P_a, \quad G_a \to c G_a, \qquad c\to \infty \,.
\end{equation}
In this limit, $\{ P_a, J,D,H\}$ form a $z=2$ Lifshitz subalgebra. 
Interesting, a $z=2$ Lifshitz spacetime solution can be indeed found in the Chern-Simons theory by gauging the extended  Schr\"odinger algebra \cite{Hartong:2016yrf}, 
\begin{equation}\label{eq:Lfishitzz=2-CS}
	\mathcal{A} = H \frac{dt}{r^2} + (P_1 -D) \frac{dr}{r} + (P_2-Z) \frac{dx}{r} \,.
\end{equation}
where $Z$ is the generators in the extended Schr\"odinger algebra \cite{Hartong:2016yrf} appearing in $[P_1,P_2]= Z$.
The solution \eqref{eq:Lfishitzz=2-CS} provides an alternative explanation of the $z=2$ Lifshitz geometry in Chern-Simons theory, overcoming the problems of the solution being degenerate in sl$(3,\mathbb{R})$ higher spin theory \cite{Lei:2015ika,Gary:2014mca}.
The solution \eqref{eq:vacuumsolution-SU12} provides a novel perspective for understanding the $z=2$ Lifshitz geometry \eqref{eq:Lfishitzz=2-CS}.
In the $1/c$ expansion \cite{Hansen:2020pqs} with the rescaling \eqref{eq:expansion-Aguage}, the $Z$ generator will appear as the $H \otimes c^{-1}$ generator, consistent with $[P_1,P_2]=H$ in su$(1,2)$ algebra.  
To achieve the $z=2$ Lifshitz geometry, we should further take the limit $\theta \to 0$ such that 	$dx_1 \approx dr$ and $dx_2 \approx r d\theta$ in this limit.
Then the solution \eqref{eq:vacuumsolution-SU12} is exactly reduced to the $z=2$ Lifshitz solution \eqref{eq:Lfishitzz=2-CS}. 

In summary, the vacuum solution of su$(1,2)\oplus$u$(1)$ Chern-Simons theory is conformal to the $\Omega$-deformed background in four dimensions. 
It also serves as the ``seed geometry" where we can start to approach the $z=2$ Lifshitz geometry in the appropriate limit.
Many interesting physics related to $z=2$ Lifshitz geometry in three dimensions can then be revisited from this novel perspective, including but not limited to singularity resolutions \cite{Lei:2015ika,Andrade:2014bsa,Gary:2014mca,Horowitz:2011gh,Harrison:2012vy,Bao:2012yt}, Lifshitz holography \cite{Taylor:2015glc, Hartong:2015wxa, Hartong:2014oma,Christensen:2013lma,Taylor:2008tg} and Lifshitz black holes \cite{Gutperle:2013oxa}.

\subsection*{Excitation solution}

In this subsection, we explore a class of nontrivial solutions to the flatness condition 
\begin{equation}
F= d\mA+ \mA \wedge \mA =0 \,,
\end{equation}
which describe excitations around the vacuum configuration \eqref{eq:vacuumsolution-SU12}.
By excitations, we refer to the introduction of dynamical functions—denoted by 
$(\mathcal{M},\mN,\cdots)$—that parameterize excitations from the vacuum.
The vacuum solution is recovered when these functions are set to zero.

Although all solutions to the flatness condition are locally pure gauge, physically distinct configurations may emerge due to boundary conditions and asymptotic symmetries. To construct such solutions explicitly, we adopt the standard method of separating the radial dependence via a group element $g(t,r,\theta)$, writing the gauge field in the form:
\begin{equation}\label{eq:Atoaby-b}
	\mathcal{A} = g^{-1} dg +g^{-1} a g
\end{equation}
where 
$a$ is a gauge field defined on the two-dimensional boundary manifold and also satisfies the flatness condition $da +a \wedge a =0$. 
By an appropriate choice of the group element $g$, we can absorb all radial dependence into it, reducing the problem to solving the flatness condition for $a$ in two dimensions. 
This approach mirrors the treatment of AdS$_3$ gravity in the Chern–Simons formulation, where the gauge fields $a$ and $\bar{a}$ are taken to depend only on boundary light-cone coordinates $x_{\pm}= t\pm x$, while the radial dependence is encoded in $g =e^{\rho L_0}$ (see for example \cite{Kraus:2005zm,Campoleoni:2010zq}). 
In our case, we similarly isolate the radial dependence by choosing:
\begin{equation}\label{eq:b-group-element}
	g=e^{-\rho (D-P_1) -\theta J}\,.
\end{equation}
With this choice, the vacuum gauge field \eqref{eq:vacuumsolution-SU12} can be rewritten in terms of the gauge field $a$ as 
\begin{equation}\label{eq:vacuum-SU12-a}
a_{\text{vac}} =H dt + \left(P_2 +J -\frac{1}{2} H \right) d\theta
\,,
\end{equation}
which manifestly satisfies the flatness condition in $(t,\theta)$ submanifold.

The goal is now to construct more general solutions for 
$a$ by turning on arbitrary functions $(\mathcal{M},\mathcal{N}, \cdots)$ that deform the vacuum configuration while preserving the flatness condition. These solutions play a role analogous to the Brown–Henneaux excitations \cite{Brown:1986nw} in AdS$_3$ gravity, where the inclusion of arbitrary functions leads to a rich asymptotic symmetry structure.
In the present context, such configurations are of particular interest in the framework of non-AdS holography, as they offer a potential route to understanding non-relativistic field theories that appear in condensed matter physics, including systems exhibiting $z=2$ Lifshitz scaling \cite{Gary:2014mca,Lei:2015ika, Taylor:2015glc,Taylor:2008tg}.

We now consider excited solutions in the background defined by \eqref{eq:vacuum-SU12-a}.
A simple excited solution takes the form:
\begin{equation}\label{eq:pertur-SU12-a}
a = \Big(H+\mathcal{N}(t,\theta) (P_2-H) + \mathcal{M} (t,\theta) P_1 \Big) dt + \left(P_2 +J -\frac{1}{2} H \right) d\theta \,.
\end{equation}
The somewhat unconventional combinations of generators, such as $P_2 +J-\frac{1}{2} H$ and $P_2 -H$ are chosen because they are eigenstates of the operator $D-P_1$ generator, which plays a special role in defining the group element $g$ in \eqref{eq:b-group-element}. 
These specific combinations can be viewed as arising from a nontrivial automorphism of the su$(1,2)$ algebra — though their precise algebraic interpretation remains somewhat obscure. 
Our goal is to determine the conditions under which the ansatz \eqref{eq:pertur-SU12-a} solves the flatness condition $da +a \wedge  a =0$. 
This requirement imposes differential constraints on the functions 
$(\mathcal{M}(t,\theta),\mathcal{N}(t,\theta))$: 
\begin{equation}
\partial_\theta \mathcal{N} + \mathcal{M} =0 , \qquad -\partial_\theta \mathcal{M} +\mathcal{N}=0\,,
\end{equation}
whose solutions are found to be:
\begin{align}
	\begin{split}
\mathcal{M}(t,\theta) &= C_1(t) \cos \theta +C_2(t) \sin\theta \\
\mathcal{N}(t,\theta) &= C_2(t) \cos \theta -C_1(t) \sin\theta \,,
	\end{split}
\end{align}
where $C_{1,2}(t)$ are free functions of time $t$. 
Substituting these expressions into the Chern–Simons ansatz $\mathcal{A} = g^{-1} a g +g^{-1} dg$ with the group element \eqref{eq:b-group-element}, we obtain the full gauge field:
\begin{align}
	\begin{split}
		\mathcal{A} 
		=&H\Big(\frac{dt}{r^2}-r\mathcal{N}(t,\theta)dt-\frac{1}{2}d\theta\Big)+P_1\Big(r\mathcal{M}(t,\theta)dt+C_1(t)dt+\cos\theta\frac{dr}{r}-\sin\theta d\theta\Big)\\
		&+P_2\Big(r\mathcal{N}(t,\theta)dt+C_2(t)dt+\sin\theta\frac{dr}{r}+\cos\theta d\theta\Big)-D\frac{dr}{r}\,.
	\end{split}
\end{align}
This configuration describes a time-dependent excitation around the vacuum background and, as we will see, encodes nontrivial information about the asymptotic structure of the theory.
The free functions $C_{1,2}(t)$  parameterize a class of non-relativistic excitations, and their interpretation may shed light on the dual field theory dynamics in the context of non-AdS holography.
We leave this to future work.

\section{Infinitely extended symmetry}\label{sec:infinitesymmetry}

The work in \cite{Chernyavsky:2019hyp} also explored the infinite extension of the extended Schr\"odinger symmetry \cite{Hartong:2016yrf}.
In order to make comparison between the extended Schr\"odinger algebra and the su$(1,2)\oplus$u$(1)$ algebra, we redefine the  generators of the latter \eqref{eq:newSU12} and the extended Schr\"odinger symmetry in the basis labeled by Fourier modes.
For the su$(1,2)\oplus$u$(1)$ algebra, we rewrite the generators as 
\begin{align}
	\begin{split}
& L_1 = K ,\quad L_0 = -\frac{1}{2} D, \quad L_{-1} = H, \quad J_0= J, \quad N_0 = N \\
& Y_{-\frac{1}{2}}^a = P_a, \quad Y_{\frac{1}{2}}^a = - G_a  \,,	\end{split}
\end{align}
such that the algebra \eqref{eq:newSU12} can be summarized as 
\begin{align}\label{eq:reformulation-SU12}
	\begin{split}
& [L_m,L_n] = (m-n) L_{m+n}	\qquad  [L_m, Y_r^a] = \left(\frac{m}{2} -r\right) Y_{m+r}^a  \\
& [L_m,J_n]	= - n J_{m+n} , \qquad  [J_m, Y_r^a] = \epsilon_{ab} Y_{m+r}^{b} \\
& [Y_r^a,Y_s^b] = \epsilon_{ab} L_{r+s} -\delta_{ab} (r-s) \left(\frac{3}{2} J_{r+s} -N_{r+s} \right)  \,.
		\end{split}
\end{align}
In this convention, $r,s$ take the values of half-integers while $m,n$ take the integer values. 
Similarly, the extended Schr\"odinger algebra is acquired by grading the sl$(2,\mathbb{R})$ subalgebra by introducing BMS generators $M_n = \epsilon L_n$ \cite{Gary:2014ppa}, where $\epsilon$ was claimed to be Grassmanian variable \cite{Gary:2014ppa,Krishnan:2013wta} \footnote{The Grassmanian condition is to guarantee $\epsilon^2=0$ such that $[M_n,M_m]=0$ and can be used to build up the non-degenerate bilinear product. 
However, this statement is similar to requiring the $\epsilon$ operator to be nilpotent, which is similar to truncate the series in a given order in the $1/c$ expansion.
}.
The extended Schr\"odinger algebra can be reformulated as
\cite{Baiguera:2024vlj, Chernyavsky:2019hyp,Kasikci:2020qsj}
\begin{eqnarray}\nonumber
	[L_m,L_n] &=& (m-n)L_{n+m},  \qquad 
	\left[L_m,Y_r^a \right] = \left(\frac{m}{2}-r \right) Y_{r+m}^a, \qquad \left[L_m,N_n\right] = -n N_{n+m} \\ \nonumber
	[L_m,J_n] &=& -n J_{m+n}, \qquad 
	\left[L_m,M_n\right] = (m-n) M_{n+m} , \qquad [J_n,M_m] = 2nN_{m+n} \\ \label{Differend}
	\left[Y_r^a,Y_s^b\right] &=& (r-s) N_{r+s} \delta_{ab} +\epsilon_{ab}M_{r+s}, \qquad 
	\left[J_n,Y_r^a\right] = \epsilon_{ab}Y_{r+n}^b  \,.
	\label{eq:eSchr-Lnbasis}
\end{eqnarray}
Notice  we also need a redefinition of the u$(1)$ generators $N_m$ to acquire the extended Schr\"odinger algebra \eqref{eq:eSchr-Lnbasis} from the su$(1,2)\oplus$u$(1)$ algebra \eqref{eq:reformulation-SU12}. 
The generators in \eqref{eq:reformulation-SU12} take only values $m,n=0,\pm 1$ and $r,s =\pm \frac{1}{2}$.  
Therefore, the commutators $[L_m,J_n]$ and $[L_m,N_n]$ seem trivial as they essentially vanish.
We display these commutators in this way for comparison with infinitely extended symmetries in the later part of this section.

The algebra \eqref{eq:eSchr-Lnbasis} can be identified as a bosonic counterpart of the centrally extended $\mN=2$ super BMS algebra. 
Its structure is as follows:
\begin{itemize}
	\item The generators $L_m$ and $M_m$ constitute the standard BMS algebra  (spanning the superrotations and supertranslations). 
	\item The remaining generators, comprising translations and Galilean boosts $Y_r^a$	play a role analogous to supercharges in the supersymmetric context. Crucially, they obey commutation relations mirroring those of fermionic generators in the parent superalgebra:
	\begin{equation}\label{eq:superalgebra-SZ}
		\{Q_r^a, Q_s^b \} = \rho_{ab} S_{r+s} +\epsilon_{ab}(r-s)Z_{r+s}, \quad \rho_{ab} = \left(
		\begin{array}{cc}
			0	& 1 \\
			1	& 0
		\end{array}
		\right)  \,.
	\end{equation}
Here the $Z_{r+s}$ is the central charges in the superalgebra \cite{Bagchi:2009ke,Mandal:2010gx}. 
\item The rotation $J$ now acts as the u$(1)$ R-symmetry of the supercharges while the $N$ is the central extension of the superalgebra \cite{Chernyavsky:2019hyp}.
\end{itemize}
The relation between the extended Schr\"odinger algebra  \eqref{eq:eSchr-Lnbasis} and the $\mN=2$ super BMS algebra can be realized by introducing two Grassmanian variables $\theta_{1,2}$ satisfying $\theta_1 \theta_2 = -\theta_2 \theta_1$ such that
\begin{equation}\label{eq:Grassmanianize}
	Y_r^1 = \theta_1 Q_r^1 +\theta_2 Q_r^2, \qquad  	Y_r^2 = -i \theta_1 Q_r^1 + i \theta_2 Q_r^2  \,.
\end{equation}
This structure automatically relates the anticommutators of the supercharges $Q_r^a$ to the commutators of $Y_r^a$ \footnote{This proposal was enlightened by Prof. Jelle Hartong.}. 
Furthermore, the Grassmannian operator $\epsilon$, which links the BMS generators via $M_n= \epsilon L_n$, is then identified as $\epsilon= \theta_1\theta_2$. 
This reflects the nilpotent nature of the supersymmetric extension and highlights the algebraic consistency of the construction.
The observation that the extended Schrödinger algebra constitutes the finite subalgebra of the $\mathcal{N}=2$ super BMS algebra raises a deeper question: How can this bosonic counterpart of the super BMS algebra be understood from the perspective of the su$(1,2)\oplus$u$(1)$ Chern-Simons theory?
This prompts an investigation into whether the su$(1,2)$ symmetry structure—particularly with its u$(1)$ extension—naturally encodes  the supersymmetric features manifested in the non-relativistic limit.

\begin{figure}
	\centering
	\includegraphics[trim=6cm 13cm 2cm 5cm,width=0.6\linewidth]{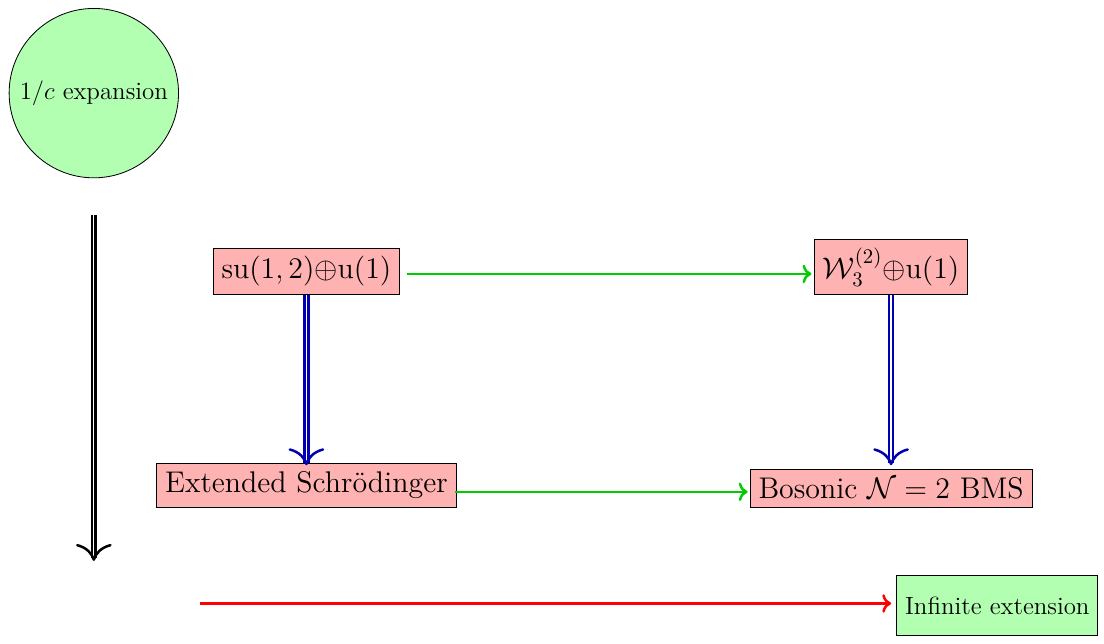}
	\caption{The algebra in seed theory and the $1/c$ expanded algebra are related in both the finite parts of the algebra and also the infinite extended parts. 
	}
	\label{fig:diagrams}
\end{figure}

In fact, the $\mathcal{W}_3^{(2)}$ algebra is known to be bosonic analogue of $\mN=2$ superconformal algebra \cite{Bilal:1991cf,Afshar:2012nk,Ammon:2011nk}, which includes SL$(2,R)$ and the R-symmetry. 
The $1/c$ expansion action on sl$(2,\mathbb{R})$ algebra exactly generate the 2d finite Galilean conformal algebra \cite{Bagchi:2009my}, while its action on Virasoro algebra generates the infinitely extended BMS$_3$ algebra.  
In this similar spirit, we can expect that the $\mathcal{W}_3^{(2)}$ with the affine u$(1)$ extension could be a potential candidate to generate the infinite extension of the extended Schr\"odinger algebra \eqref{eq:eSchr-Lnbasis} after performing the $(1/c)$-expansion. 
A salient conflict in this expectation is, that the $\mathcal{W}_3^{(2)}$ algebra is nonlinear while the bosonic super BMS$_3$ algebra is linear \cite{Chernyavsky:2019hyp}. 
Then, one would wonder if one starts with an solution in seed theory -- su$(1,2)\oplus$u$(1)$ Chern-Simons theory, whose $(1/c)$-expansion can give rise to the background generating the linear bosonic super BMS$_3$ algebra, then what is the infinitely extended symmetry of this background in the seed theory?
Especially one would wonder whether the algebra is nonlinear. 
As we show below, the su$(1,2)\oplus$u$(1)$ Chern-Simons theory indeed has a solution to the flatness condition whose $(1/c)$-expansion can give rise to the flat spacetime BMS solution.
This solution would enjoy an infinitely extended symmetry to be $\mathcal{W}_3^{(2)} \oplus$u$(1)$ affine, as dictated in the figure \ref{fig:diagrams}.  

As in AdS$_3$ and flat geometry \cite{Campoleoni:2010zq,Chernyavsky:2019hyp}, we consider the solution to be in the following gauge
\begin{equation}
	A = b^{-1} db + b^{-1} a b \,,
\end{equation}
such that $b$ depends on the radial direction. 
In this decomposition, the gauge field $a$ only depends on two spacetime coordinates, denoted as $(t,x)$. 
We also take $x^\pm = t\pm x$.
We thus take the non-degenerate solution as 
\begin{equation}\label{eq:solution-SL2-SU12}
a  = (L_1 + \mathcal{L} L_{-1} + \alpha J +\mathcal{W}_a Y_{-\frac{1}{2}}^a + \beta N) dx^+  \,.
\end{equation}
To solve the flatness condition, the functions $\mathcal{L},\mW_a,\alpha,\beta$ are all functions of $x^+$.
By performing a $(1/c)$-expansion and redefining the time coordinate via $c \tilde{t} = t$, the symmetry generators are rescaled according to \eqref{eq:expansion-Aguage}. 
When this rescaling is combined with the Sugawara shift redefinition
\begin{equation}
\mathcal{M } \to \mathcal{M } - \mathcal{N}^2, \qquad \mathcal{L} \to \mathcal{L}- \mN \mathcal{I}\,,
\end{equation} 
the solution \eqref{eq:solution-SL2-SU12}  correctly reproduces the corresponding solutions in the reformulated bosonic supergravity theory \cite{Chernyavsky:2019hyp}: 
\begin{align}\label{eq:sol-asym-flat}
	\begin{split}
a_\phi &= L_1 +(\mathcal{N}^2-\mathcal{M}) L_{-1} +(\mathcal{N}  \mathcal{I} -\mathcal{L}) M_{-1} +\sqrt{2} \mathcal{C}^a Z^a_{-\frac{1}{2}} + \mathcal{I} N+ \mathcal{N} I \\
a_{\tilde{t}} &= M_1 +(\mathcal{N}^2-\mathcal{M}) M_{-1} +2\mathcal{N} N   \,.
	\end{split}
\end{align}
These solutions exhibit structural parallels with those found in asymptotically flat supergravity theories. Consequently, the solution presented in \eqref{eq:solution-SL2-SU12} emerges as the prime candidate for investigating the infinite-dimensional symmetry extension within the su$(1,2)\oplus$u$(1)$ Chern-Simons framework. 

Now consider the variation of gauge field $a$ solution \eqref{eq:solution-SL2-SU12} parametrized as
\begin{equation}\label{eq:trans-equation}
	\delta a = d\epsilon + [a,\epsilon]\,,
\end{equation}
where the transformation parameter is explicitly
\begin{equation}
\epsilon = \epsilon^{L_m} L_m + \epsilon_J J + \epsilon_{N} N + \epsilon^{Y_{r}^a} Y_{r}^a \,,
\end{equation}
for $m=\pm 1, 0$ and $r= \pm \frac{1}{2}$. We denote the variation parameters as 
\begin{equation}
\epsilon^{L_1} \equiv \epsilon_L, \qquad \epsilon^{Y_{\frac{1}{2}}^a} \equiv \epsilon_{a}\,,
\end{equation}
which solve the transformation equation \eqref{eq:trans-equation} as
\begin{align}
\begin{split}
\delta \mL & = \mL' \epsilon_L+2\mL \epsilon_L'- \frac{1}{2} \epsilon_L''' -\frac{3}{2} \mW_2 \epsilon_1' - \left(\frac{1}{2} \mW_2'-\alpha \mW_1 \right) \epsilon_1 \\
& +\frac{3}{2} \mW_1 \epsilon_2' +\left(\frac{1}{2} \mW_1' +\alpha \mW_2 \right) \epsilon_2 \\
\delta \mW_1 & = 2\alpha\epsilon_2' +\alpha' \epsilon_2 + \mW_2 \epsilon_J  + (\mL + \alpha^2) \epsilon_1 -\epsilon_1''  + (\mW_1'-\alpha \mW_2) \epsilon_L + \frac{3}{2} \mW_1 \epsilon_L'
\end{split} \\
\delta \mW_2 & = -2\alpha \epsilon_1'-\alpha' \epsilon_1 -\mW_1 \epsilon_J + (\mL +\alpha^2) \epsilon_2 -\epsilon_2'' +(\mW_2'+\alpha \mW_1) \epsilon_L + \frac{3}{2} \mW_2 \epsilon_L' \\
\delta \alpha & = \epsilon_J' +\frac{3}{2}\mW_1 \epsilon_1 + \frac{3}{2} \mW_2 \epsilon_2\, \qquad
\delta \beta  = \epsilon_J' - \mW_1 \epsilon_1 - \mW_2 \epsilon_2\,.
\end{align}
The corresponding conserved charges are determined by the bilinear product as
The charge is 
\begin{equation}\label{eq:SU12-charge}
	Q=\int dx  \left[\epsilon_L(c_S \mathcal{L})-\epsilon_2(c_S \mathcal{W}_1)+\epsilon_1(c_S \mathcal{W}_2)+\epsilon_J \left(\frac{2}{3}(c_S+c_U)\alpha+c_U\beta \right)+ \left( c_U\alpha + \frac{3}{2} \beta c_U \right) \epsilon_N \right].
\end{equation}
For convenience, we redefine the conserved functions up to the Sugawara shift
\begin{align}
	\begin{split}
&	\tilde{\beta}  = \frac{3}{2} \beta +\alpha, \quad \tilde{\alpha} = \frac{2}{3} c_S \alpha + \frac{2}{3} c_U \tilde{\beta}\,, \quad  \tilde{\mathcal{L}} = \mathcal{L} + \frac{1}{3c_S^2} \left( \frac{3}{2} \tilde{\alpha} -c_U \tilde{\beta} \right)^2\,, 
	\end{split}
\end{align}
This redefinition will remove the nonlinear terms in $\{\mathcal{L},\mathcal{W}_a\}$ such that the Poisson brackets are 
\begin{align}
	\begin{split}
		\{ \tilde{\mathcal{L}}(x),\tilde{\mathcal{L}}(y)\} &= \tilde{\mathcal{L}}'(x) \delta(x-y) + 2\tilde{\mathcal{L}}(x) \delta'(x-y) +\frac{1}{2} \delta'''(x-y) \\
		\{\tilde{\mathcal{L}} (x),\mathcal{W}_a(y)\} &= \frac{3}{2} \mathcal{W}_a(x) \delta'(x-y) + 
		\frac{1}{2}\mathcal{W}_a'(x) 
		\delta(x-y) \\
		\{ \tilde{\mathcal{L}}(x),\tilde{\alpha}(y)\} &=   \left[\tilde{\alpha}(x) -\frac{2}{3}c_U \tilde{\beta}(x) \right] \delta'(x-y)\,,
		\qquad \{ \mathcal{L}(x),\tilde{\beta}(y)\} = \{ \mathcal{W}_a(x),\tilde{\beta}(y)\}=0  \\
		\{ \mathcal{W}_a(x),\mathcal{W}_a(y)\} &= - \left( \frac{3}{2} \tilde{\alpha}'(x) -c_U \tilde{\beta}'(x)  \right) \delta(x-y) -2 \delta'(x-y) \left( \frac{3}{2} \tilde{\alpha}(x) -c_U \tilde{\beta}(x)  \right) \\
		\{ \mathcal{W}_2(x),\mathcal{W}_1(y)\} &= \left[\tilde{\mathcal{L}} (x)- \frac{4}{3c_S^2}\left( \frac{3}{2} \tilde{\alpha}(x) -c_U \tilde{\beta}(x) \right)^2 \right] \delta(x-y) +\delta''(x-y) \\
		\{\mathcal{W}_a(x),\tilde{\alpha}(y) \} &=  \epsilon_{ab} \mathcal{W}_b(x) \delta(x-y)\\
		\{\tilde{\alpha}(x), \tilde{\alpha}(y) \} & = \frac{2}{3}(c_U+c_S) \delta'(x-y), \quad  \{\tilde{\alpha}(x), \tilde{\beta}(y) \}  = \delta'(x-y) \\
		\{\tilde{\beta}(x), \tilde{\beta}(y) \} & = \frac{3}{2c_U} \delta'(x-y) \,.
	\end{split}
\end{align}
We can then define the Fourier modes as follows
\begin{align}
	\begin{split}
		L_n &= \int dx  \tilde{\mathcal{L}}(x) e^{in x}, \quad Y_r^{a} = \frac{1}{\sqrt{-i}} \int dx \mathcal{W}_a(x) e^{i r x} \\
		J_n & = i \int dx \tilde{\alpha}(x) e^{in x} ,\qquad N_n  = i c_U \int dx \tilde{\beta}(x) e^{in x} \,.
	\end{split}
\end{align}
The corresponding algebra thus has the following structure (with shift $L_0 \to L_0 + \frac{\pi}{2} c_S$ in the zero mode): 
\begin{align}\label{eq:infinitesymmetry-W32algebra}
	\begin{split}
		[L_m ,L_n]& = (m-n) L_{m+n} + c_S \pi (n^3-n) \delta_{n+m,0}, \qquad [L_m,N_n] = 0 \\
		[L_m ,Y_r^a]&= \left(\frac{m}{2}-r \right) Y_{m+r}^a\,,\quad  [L_m, J_n] = -n \left( J_{n+m} -\frac{2}{3} N_{n+m} \right) \\
		[Y_r^a,Y_s^a] &= -(r-s) \left(\frac{3}{2} J_{r+s} - N_{r+s} \right) \\
		[Y_r^1,Y_s^2] &=L_{r+s} + \frac{2}{3\pi c_S} \left(
		\frac{3}{2} J_{r+s-k} -N_{r+s-k}
		\right) \left( \frac{3}{2} J_k - N_k \right) -2\pi c_S \left(r^2-\frac{1}{4} \right)\delta_{r+s,0} \\
		[J_n ,Y_r^a] &= \epsilon_{ab} Y_{n+r}^b, \quad [J_m,J_n] = \frac{4\pi }{3}n (c_U+c_S) \delta_{n+m,0}, \\
		[J_m ,N_n] &= 2\pi c_U n \delta_{n+m,0} , \qquad [N_m,N_n] = 3\pi c_U  n \delta_{n+m,0} \,.
	\end{split}
\end{align}
The algebra \eqref{eq:infinitesymmetry-W32algebra} includes the Virasoro algebra generated by the modes $L_n$, and corresponds precisely to the so-called $\mathcal{W}_3^{(2)}$ algebra, equipped with an affine u$(1)$ extension parametrized by the modes $N_m$ \cite{Afshar:2013vka}.
The generators $L_n$ span the standard Virasoro algebra, while $J_n$ correspond to a u$(1)$ Kac-Moody current.
The generators $Y_r^a$ can be interpreted as bosonic analogues of supercharges in the $\mathcal{N}=2$ superconformal algebra, an analogy that becomes manifest upon introducing the Grassmannian extension described in \eqref{eq:Grassmanianize}.
The parameters $c_S$ and $c_U$, originating from the non-degenerate bilinear form, appear as central charges in this infinite-dimensional symmetry algebra.
Importantly, unlike the linear $\mathcal{N}=2$ superconformal algebra (see e.g., \cite{Ammon:2011nk}), the algebra presented here is intrinsically non-linear: the commutator $[Y^1_r, Y^2_s]$ contains bilinear combinations of $J_n$ and $N_n$.
This non-linearity is characteristic of $\mathcal{W}$-type algebras and plays a crucial role in encoding the asymptotic symmetry structure of higher-spin or non-Riemannian gravitational theories.

In our construction, this extended algebra emerges as the asymptotic symmetry algebra associated with the background solution \eqref{eq:solution-SL2-SU12}.
Performing a $(1/c)$-expansion of this algebra and organizing the higher-order generators as in \eqref{eq:expansion-Aguage} yields the infinite-dimensional extension of the Schrödinger algebra discussed in \cite{Chernyavsky:2019hyp}.
Within this expansion, the non-linear terms are relegated to subleading orders and thus become suppressed, resulting in a linear algebraic structure at leading order.
While this suggests a rich underlying symmetry structure, it remains an open question whether the full non-linear algebra governs the asymptotic symmetries of the torsional Newton-Cartan background \eqref{eq:vacuumsolution-SU12}.
We intend to explore this intriguing possibility in future work.

\section{Discussion}\label{sec:discussion}

In this paper, we discuss the TNC gravity theory built up by gauging the su$(1,2)\oplus$u$(1)$ algebra and construct an action for the TNC gravity theory using the Chern-Simons theory.
The theory has a general form of torsion and can be related to the Schr\"odinger gravity by $(1/c)$-expansion \cite{Hansen:2020pqs}.
The three independent parameters of Schr\"odinger gravity correspond to the leading order constraints and subleading order dynamics in $(1/c)$-expansion, respectively. 
This resolves the long-standing puzzles in searching for the seed theory of the Schr\"odinger gravity by either $(1/c)$-expansion or null reduction.
Besides, we also explain the identifications between the extended Schr\"odinger algebra \cite{Hartong:2016yrf} and bosonic resembles of super BMS algebra.
This identification was used to reformulate the Chern-Simons theory in extended Schr\"odinger algebra as an asymptotically flat supergravity theory \cite{Chernyavsky:2019hyp}, which is regarded as the alternative vacuum of the Chern-Simons theory. 
We treat the relation between two algebras as the consequence of $1/c$ expansion of $\mW_3^{(2)}\oplus$u$(1)$ algebra (as the infinite extension of su$(1,2)\oplus$u$(1)$ algebra), which acts as the bosonic analogue of superconformal algebra.
We hope to revisit this problem and scan the Chern-Simons theory via different vacua in the future. 

As a further development, the alternative choice of vacuum of the Chern-Simons theory is related to different bases to represent the su$(1,2)$ algebra, which are adapted to different models \cite{Bars:1989bb,Gunaydin:2007qq}. 
For example, the Harish-Chandra coordinate basis of the su$(1,2)$ algebra is similar to the harmonic oscillator representations.
The basis considered in this paper is analogous to the diagonal embedding of sl$(2)$ in su$(1,2)$ \cite{Chernyavsky:2019hyp,Campoleoni:2010zq,Ammon:2011nk}. 
An alternative basis can be written down by taking the sets of harmonic oscillators as ($\alpha,\beta=1,2$) \cite{Harmark:2007px}
\begin{equation}
	[\mba_\alpha,\mba^\dagger_\beta] = \delta_{\alpha \beta}, \qquad [\mbb,\mbb^\dagger] =1\,,
\end{equation}
such that the  algebra generators are 
\begin{eqnarray} \label{eq:newbasisforSU21-1}
L_0 &=&\frac{1}{2} (1+\mba_1^\dagger \mba_1 +\mbb^\dagger \mbb), \quad L_1= \mba_1^\dagger \mbb^\dagger, \quad L_{-1} =\mba_1 \mbb \\  \label{eq:newbasisforSU21-2}
	\tilde{L}_0 &=&   \frac{1}{2} (1+\mba_2^\dagger \mba_2 +\mbb^\dagger \mbb), \quad \tilde{L}_1 = \mba_2^\dagger \mbb^\dagger, \quad \tilde{L}_{-1}=\mba_2 \mbb \\ \label{eq:newbasisforSU21-3}
	J_+ &=& \mba_1^\dagger \mba_2, \qquad J_- =\mba_2^\dagger \mba_1
	\,.
\end{eqnarray}
In this basis, the su$(1,2)$ algebra can be viewed as two coupled su$(1,1)$ algebras \eqref{eq:newbasisforSU21-1} and \eqref{eq:newbasisforSU21-2}. 
This basis is quite convenient for studying states in Spin Matrix theory, where the derivative letters are generated by the numbers of $L_{1}$ and $\tilde{L}_{1}$ generators, which act on the primary operators in $\mN=4$ SYM \cite{Baiguera:2020mgk, Baiguera:2022pll}. 
The automorphism between two basis \eqref{eq:newbasisforSU21-1}-\eqref{eq:newbasisforSU21-3} and \eqref{eq:newSU12} is given by \cite{Baiguera:2024vlj}
\begin{align}
	\begin{split}
		& iJ = \frac{2}{3} (L_0 + \tilde{L}_0) , \qquad D= 2(L_0 -\tilde{L}_0), \quad K= -2 J_+, \quad H = \frac{1}{2} J_- \\
		& G_1+iG_2 =2 L_1, \quad P_1-i P_2 = i L_{-1}, \quad P_1+iP_2 =  \tilde{L}_1, \quad G_1-iG_2 = -2i \tilde{L}_{-1}\,.
	\end{split}
\end{align}
Therefore it is interesting to understand what could be the gravity theory related to the su$(1,2)\oplus$u$(1)$ Chern-Simons theory in the basis \eqref{eq:newbasisforSU21-1}-\eqref{eq:newbasisforSU21-3}. 

A natural extension of the current framework involves constructing Newton-Cartan gravity through the gauging of the su$(1,n)\oplus$u$(1)$ group, which governs the spacetime geometry in odd dimensions $d=2n-1$. 
This group structure emerges from the null reduction procedures applied to conformal field theories in $d=2n$ dimensions, where the underlying conformal symmetry is characterized by the so$(2,2n)$ group \cite{Lambert:2021nol}.
The minimal non-trivial extension would be $n=3$, yielding a five-dimensional Newton-Cartan manifold.  
Crucially, the profound connections between such geometries and M-theory dynamics have been systematically explored in previous works \cite{Lambert:2019jwi, Lambert:2020zdc,Lambert:2020scy}.
The geometric framework becomes tangible through its realization in the su$(1,3)$ algebra, 
whose technical details will be reviewed in Appendix \ref{appendix:SU13}. 
As evidenced by the non-vanishing commutator of spatial translations \eqref{eq:SU13algebra}, this algebraic structure exhibits non-trivial characteristics that depart from conventional relativistic spacetime symmetries: 
\begin{equation}
	[P_i,P_j] = \frac{1}{R} \Omega_{ij} H, \qquad 	\Omega_{ij} = \left( 
	\begin{array}{cc}
		0	& \mathbbm{1}_2 \\
		-\mathbbm{1}_2 	& 0
	\end{array}
	\right), \qquad i=1,...,4\,,
\end{equation}
where $\Omega_{ij}$ is the complex structure matrix.
This results in a curvature tensor related to torsion: 
\begin{equation}
	R(H)= dE^0 -2B \wedge E^0 +E^1 \wedge E^3 +E^2 \wedge E^4\,,
\end{equation}
yielding a torsional Newton-Cartan geometry as $E^0 \wedge dE^0 \neq 0$. 
Besides, the non-trivial dependence of $D,J^\alpha$ generators in the $[P_i,G_j]$ commutator again indicates that the corresponding gauge fields are not invariant under the Galilean boost transformation. 
Again, these are not in the type of torsional Newton-Cartan geometry reported in \cite{Bergshoeff:2014uea,Hartong:2015zia}. 
The corresponding affine connection will also encounter a conflict between boost invariance and dilatation invariance, as we discussed in Section \ref{ssec:gaugealgebra}.
We will leave the study of these TNC geometries to future work. 

Our work establishes the su$(1,2)\oplus$u$(1)$ as the fundamental ``seed algebra" underlying the extended Schr\"odinger algebra.
A significant problem involves determining an analogous seed algebra for the two-dimensional Galilean conformal algebra (GCA$_{d=2}$) \cite{Bagchi:2009my}, where $d$ denotes the spatial dimension.
Although GCA$_{d=2}$ constitutes a similar conformal extension of the Galilean algebra, it features a more complex structure incorporating additional special conformal generators that distinguish it from the Schr\"odinger case. 
Its structural organization bears a closer resemblance to so$(2,3)$ algebra expressed in Galilean basis, as detailed in appendix \ref{appendix:SO32} \footnote{For instance both algebra have three special conformal transformation generators, but still differ by the dynamical exponent $z$ of $H$ generator}. 
The construction of a gauge theory based on GCA$_{d=2}$ (with potential extensions) could provide a pathway to develop alternative three-dimensional conformally extended Newton-Cartan gravity. 
However, two significant challenges remain. 
First, the degenerate bilinear form of GCA$_{d=2}$ renders the conventional Chern-Simons theories formulated for this algebra pathological. 
Second, even if we can follow the extension procedures like $z=2$ Schr\"odinger algebra with two spatial directions \cite{Hartong:2016yrf}, multiple extension pathways exist without clear selection criteria, which courses combinatorial complexity.
A possible resolution is aligned with our current methodology, which would first identify the fundamental seed algebra, then systematically generate (extended) GCA$_{d=2}$ by $(1/c)$-expansions \cite{Hansen:2020pqs} \footnote{Possibly this seed algebra is exactly the so$(2,3)$ but more work is needed.}. 
Successful implementation of this paradigm would enable the development of novel non-relativistic gravitational theories via Chern-Simons formulations. 

The similarity between so$(2,3)$ and su$(1,2)\oplus$u$(1)$ Chern-Simons theories—after fixing the curvature constraints—suggests that techniques developed for the former could be applied to study the latter, particularly in holographic contexts. 
Notably, the vacuum solutions of so$(2,3)$ Chern-Simons theory include the AdS$_3$ geometry, enabling the study of gravitational Chern-Simons holography even without an Einstein-Hilbert term. 
The theory admits three distinct boundary conditions, each potentially supporting different boundary dynamics \cite{Afshar:2011qw,Afshar:2011yh}. 
Besides, the work \cite{Fuentealba:2020zkf} found alternative interesting boundary condition to accommodate black hole solutions found in \cite{Oliva:2009hz}. 
In contrast, the vacuum solution of the Chern-Simons theory features torsional geometry \eqref{eq:vacuumsolution-SU12}, which connects to both the $z=2$ Lifshitz solution by $(1/c)$-expansion and 4d $\Omega$-deformed background by null reduction and conformal mapping (shown in Section \ref{ssec:action}).
Thus, it serves as the background dual to the ground state of the non-relativistic dual field theory, offering a promising avenue to explore holography beyond relativistic regimes. 

There are many condensed matter systems with Lifshitz symmetry for $z\neq 2$ whose strongly coupled limit might be holographically dual to non-relativistic geometries with the same dynamical exponent. 
Efforts to construct anisotropic geometries with integer dynamical exponent $z$ in sl$(z+1,\mathbb{R})$ Chern-Simons theory was done in \cite{Gary:2012ms}.
However, as emphasized in this paper, the solutions in these gauge theories were shown not equivalent to Lifshitz gravity due to the degenerate nature of the gauge theory \cite{Lei:2015ika}. 
Our current work together with \cite{Hartong:2016yrf} explained that the su$(1,2)\oplus$u$(1)$ Chern-Simons theory with its vacuum solution is the appropriate framework for constructing the $z=2$ Lifshitz spacetime. 
And the seed geometry to generate the $z=2$ Lifshitz spacetime is the null reduction of a geometry conformally equivalent to 4d $\Omega$ background. 
The corresponding open question is, are there other geometries serving as the seed geometry for Lifshitz spacetimes with $z\neq 2$? 
This is also related to understanding the appearance of the integer $z\neq 2$ Lifshitz solution in sl$(z+1,\mathbb{R})$ Chern-Simons by Newton-Cartan theories. 
There might be two ways to address this problem.
On one hand, we may explore whether a suitable larger symmetry algebra—analogous to su$(1,n)$ , which serves as the extended symmetry for the $z=2$ Lifshitz case — can be constructed for the $z\neq 2$ Schr\"odinger algebra. 
It is not clear whether the unknown algebra can contain a special conformal transformation, except such special conformal symmetry was proposed for $z\neq 2$ Schr\"odinger algebra in \cite{Hartong:2014pma,Hartong:2014oma}.
On the other hand, we can also examine whether the Lifshitz geometries in sl$(z+1,\mathbb{R})$ Chern-Simons theory can resolve the degeneracy problems by introducing extra U$(1)$ gauge fields, as requested by Newton-Cartan geometry. 
This might need the studies of the sl$(z+1,\mathbb{R})$ or the su$(1,z)$ algebra in a novel basis. 
Notice here the su$(1,z)$ algebra needs to be organized in the basis not adapted to $z=2$ Lifshitz scaling but Lifshitz scaling with general integer $z$ \cite{Gary:2012ms}. 

Our current work represents significant progress towards identifying seed theories that can generate interesting Newton-Cartan gravity through either null reduction or $1/c$ expansion.
The Chern-Simons theory studied in this paper still describes non-relativistic torsional Newton-Cartan gravity in nature. 
This suggests that it may emerge from the null reduction of a four-dimensional relativistic gravitational theory, which is supported by several key observations \cite{Lambert:2021nol}.
First, su$(1,2)\oplus$u$(1)$ naturally arises from the four-dimensional conformal symmetry group so$(2,4)$.
Second, the vacuum solution of the su$(1,2)\oplus$u$(1)$ Chern-Simons theory is conformally related to the 4d $\Omega$-deformed background \cite{Lambert:2021nol}.
These connections strongly indicate the existence of an underlying four-dimensional gravitational theory whose properties warrant further investigation.
Among potential candidates for this seed theory, four-dimensional conformal gravity \cite{Maldacena:2011mk,Grumiller:2013mxa}.
appears particularly promising. 

Finally, equipped with the four-dimensional Lorentzian seed theory, we can systematically explore the ultra-relativistic (Carrollian) limit of these gravitational theories, which acts as a complementary direction to investigate by taking the speed of light to zero. 
This limit has garnered increasing attention in recent years, particularly due to its relevance in flat space holography and null boundary dynamics \cite{Hartong:2015usd,Hartong:2015xda, Adami:2023wbe,Bagchi:2023cen,Pekar:2024ukc,Hartong:2025jpp,Chen:2023pqf,Donnay:2022wvx}. 
In this framework, Carrollian geometry replaces Newton-Cartan as the underlying spacetime structure, and various consistent Carrollian gravity theories have been constructed \cite{Concha:2024tcu,Bergshoeff:2024ilz, Hansen:2021fxi,Musaeus:2023oyp, March:2024zck, Pekar:2024ukc, Ecker:2023uwm,deBoer:2023fnj, Concha:2022muu, Campoleoni:2022ebj, Figueroa-OFarrill:2022mcy,Ecker:2025ncp}. 

The four-dimensional seed theories proposed in this paper may admit a well-defined Carrollian limit, either via null reduction or through a $c \to 0$ ultra-relativistic expansion \cite{Hansen:2021fxi}.
A natural question is whether the resulting three-dimensional Carrollian gravity can be formulated as a Chern-Simons theory, potentially exhibiting features of conformal gravity.
Such an investigation could provide valuable insights into a unified Chern-Simons framework that encompasses both non-relativistic and ultra-relativistic limits, and may also shed light on corresponding holographic descriptions, offering a better understanding of the dual field theories.
We leave a detailed exploration of this direction to future work.

\section*{Acknowledgement}

We thank Stefano Baiguera, Oscar Fuentealba, Jelle Hartong, Gerben Oling and Ziqi Yan for useful discussions and also anonymous referee for useful comments. 
Y.L. is supported by a Project Funded by the Priority Academic Program Development of Jiangsu Higher Education Institutions (PAPD) and by National Natural Science Foundation of China (NSFC) No.12305081 and the international exchange grant between NSFC and Royal Society No.W2421035.

\appendix 

\section{Gauging algebra and Newton-Cartan gravity}
\label{sec:review-NC-gauge}

Newton–Cartan gravity can be systematically constructed by gauging non-relativistic symmetry algebras.
Gauging the Bargmann algebra leads to torsionless Newton–Cartan gravity \cite{Andringa:2010it}, while gauging the Schr\"odinger algebra yields the so-called twistless torsional Newton–Cartan (TTNC) gravity \cite{Bergshoeff:2014uea,Hartong:2016yrf}.
In this appendix, we briefly review both constructions: torsionless Newton–Cartan gravity via gauging the Bargmann algebra, and twistless torsional Newton–Cartan gravity via gauging the Schr\"odinger algebra.
For a more comprehensive treatment of these approaches, we refer the reader to \cite{Hartong:2015zia,Hartong:2016yrf}.

\subsection{Gauging Bargmann}

The Bargmann algebra consists of the time translation generator $H$, spatial translation $P_a$, rotation $J_{ab}$, Galilean boost $G_a$ and central extension $N$. 
We are interested in a theory defined on a three-dimensional manifold, so the spatial index runs over $a=1,2$. 
The non-vanishing commutation relations are given by $(J_{ab}=J \epsilon_{ab})$
\begin{align}
	\begin{split}
& [H,G_a] =P_a, \quad [P_a ,G_b ] =N, \quad [J,P_a] =\epsilon_{ab} P_b, \qquad [J,G_a] =\epsilon_{ab} G_b \,.
	\end{split}
	\end{align}
The corresponding gauge field takes the form
\begin{equation}
	\mathcal{A}_\mu = H \tau_\mu +P_a e^a_\mu + G_a \omega^a_\mu + J \omega_\mu+ N m_\mu \,.
\end{equation}
The $R(H)$ component of the strength tensor $F=d\mathcal{A} +\mathcal{A} \wedge \mA$ is $R(H)=d\tau$. 
Imposing this curvature constraint 
$R(H)=0$ corresponds to a vanishing temporal torsion condition, and leads to torsionless Newton–Cartan gravity.

The gauge field transforms under an internal Bargmann gauge transformation $\Sigma = G_a \lambda^a + J \lambda + N \sigma$ as well as under a spacetime diffeomorphism generated by a vector field $\xi^\mu$ , according to the combined transformation rule
\begin{equation}\label{eq:Gauge-transform-A}
	\bar{\delta} \mathcal{A}_\mu  = \mL_\xi \mA_\mu + \partial_\mu \Sigma + [\mathcal{A}_\mu ,\Sigma]\,.
\end{equation}
where $\mL_\xi$ denotes the Lie derivative along $\xi^\mu$.
This expression captures both the diffeomorphism covariance and the structure of the gauge algebra.
The transformation properties of the component fields in the Newton–Cartan geometry can then be extracted from the above relation. Explicitly, the fields transform as follows:
\begin{align}\label{eq:gauge-bargmmann}
	\begin{split}
\bar{\delta} \tau_\mu &= \mL_\xi \tau_{\mu} \\
\bar{\delta} e_\mu^a &= \mL_\xi e_\mu^a + \lambda \epsilon_{ab} e_\mu^b + \lambda^a \tau_\mu \\
\bar{\delta} \omega_\mu^a &= \mL_\xi \omega_\mu^a +\partial_\mu \lambda^a +\lambda \epsilon_{ab}  \omega_\mu^b + \lambda^b \omega_{\mu} \epsilon_{ab}\\
\bar{\delta} \omega_\mu &=  \mL_\xi \omega_\mu +\partial_\mu \lambda\\
\bar{\delta} m_\mu &=\mL_\xi m_\mu +\partial_\mu \sigma +e_\mu^a \lambda_a \,.
	\end{split}
\end{align}

To define covariant derivatives compatible with Newton–Cartan geometry, we introduce an affine connection $\Gamma_{\mu\nu}^\rho$.
The covariant derivatives of the temporal and spatial vielbeins are defined as:
\begin{align}\label{eq:covariant-Torless-NC}
	\begin{split}
\mathcal{D}_\mu \tau_\nu & = \partial_\mu \tau_\nu -\Gamma_{\mu \nu}^{\rho} \tau_\rho \,, \\
\mathcal{D}_\mu e_\nu^a & =\partial_\mu e_\nu^a   -\Gamma_{\mu \nu}^{\rho} e_\rho^a -\omega_\mu^a \tau_\nu -\omega_\mu \epsilon_{ab} e_\nu^b  \,.
	\end{split}
\end{align}
The affine connection transforms under spacetime diffeomorphisms and behaves as a tensor under internal transformations:
\begin{equation}
\bar{\delta} \Gamma_{\mu\nu}^\rho = \partial_\mu \partial_\nu\xi^\rho + \xi^\sigma \partial_\sigma \Gamma_{\mu\nu}^\rho + \Gamma_{\sigma \nu}^\rho \partial_\mu \xi^\sigma + \Gamma_{\mu\sigma}^\rho   \partial_\nu \xi_\sigma - \Gamma_{\mu\nu}^\sigma \partial_\sigma \xi^\rho
\end{equation}
Importantly, in the torsionless Newton–Cartan formulation, the affine connection is invariant under all internal gauge transformations, including boosts, rotations, and the central extension. This reflects the fact that the connection is entirely determined by the geometric structure encoded in $\tau_\mu,e_\mu^a$ , up to diffeomorphism covariance.

The affine connection can be determined by imposing the vielbein postulates, which ensure compatibility between the covariant derivative and the Newton–Cartan geometric structure:
\begin{equation}\label{eq:vielbein-postulate-TT}
\mathcal{D}_\mu \tau_\nu = \mathcal{D}_\mu e_\nu^a=0 \,.
\end{equation}
Solving these conditions yields the expression for the connection:
\begin{equation}\label{eq:affine-torsionless-frame}
	\Gamma_{\mu\nu}^\rho = -v^\rho \partial_\mu\tau_\nu + e^\rho_a (\partial_\mu e_\nu^a -\omega_\mu^a \tau_\nu -\omega_\mu \epsilon^{ab} e_\nu^b)\,,
\end{equation}
where the inverse vielbeins $v^\mu$ and $e^\rho_a$ satisfy the following orthogonality and completeness relations:
\begin{equation}
	v^\mu \tau_\mu =-1, \quad v^\mu e_\mu^a =0, \quad e^\mu_a \tau_\mu =0, \quad e_\mu^b e^\mu_a =\delta_a^b \,.
\end{equation}
The connection \eqref{eq:affine-torsionless-frame} is manifestly invariant under local Galilean boosts $G_a$ and the spatial rotation $J$, reflecting the internal symmetry structure of torsionless Newton–Cartan geometry.
Metric like fields $\tau_\mu$ and $h^{\mu\nu} = e^\mu_a e^\nu_b \delta_{ab}$ are invariant under the Galilean boosts $G_a$ and  rotation $J$. 
To construct quantities that are additionally invariant under the local U$(1)$ gauge transformations associated with the central extension, one introduces the modified, boost-invariant combinations:
\begin{equation}\label{eq:def-hatv-barh}
\hat{v}^\mu = v^\mu -h^{\mu\nu} m_\nu	, \qquad \bar{h}_{\mu\nu} = h_{\mu\nu} -\tau_\mu m_\nu - \tau_\nu m_\mu
\end{equation}
An alternative expression for the affine connection—constructed entirely from the metric-like fields and explicitly invariant under local Galilean boosts, spatial rotations, and U$(1)$ gauge transformations—is given by \cite{Hartong:2015zia,Bergshoeff:2014uea}
\begin{equation}
	\Gamma_{\mu\nu}^\rho = -\hat{v}^\rho \partial_\mu \tau_\nu + \frac{1}{2} h^{\rho \sigma} (\partial_\mu \bar{h}_{\sigma \nu} + \partial_\nu \bar{h}_{\mu\sigma} -\partial_\sigma \bar{h}_{\mu\nu})
\end{equation}
This form is particularly useful when studying Newton–Cartan gravity coupled to matter fields, as it makes all local symmetries manifest.

\subsection{Gauging $z=2$ Schr\"odinger algebra}
In this section, we briefly review the procedure for gauging the Schr\"odinger algebra in three spacetime dimensions with dynamical exponent $z=2$. 
For the generalization to arbitrary dynamical exponent $z \neq 2$, we refer the reader to \cite{Bergshoeff:2014uea}.
The three-dimensional Schr\"odinger algebra with $z=2$ includes time translation $H$, spatial translation $P_a$, Galilean boosts $G_a$, rotation $J$, dilatation $D$, special conformal transformation $K$ and the central extension $N$. 
The non-vanishing commutation relations are:
\begin{align}\label{eq:Schroedinger-algebra}
	\begin{split}
		& [J,P_a] =\epsilon_{ab} P_b, \qquad [J,G_a] =\epsilon_{ab} G_b, \quad [P_a,G_b] = N  \\
		&[H,K]=D, \quad  \quad [D,K]=2K, \quad [H,D]=2H \\
		& [K,P_a] =-G_a, \quad [H,G_a] =P_a, \quad [P_a,D]=P_a,\quad [G_a,D]=-G_a \,.
	\end{split}
\end{align}
The corresponding gauge connection $\mathcal{A}_\mu$ is constructed as a one-form valued in the Schr\"odinger algebra
\begin{equation}
	\mathcal{A}_\mu = H \tau_\mu +P_a e^a_\mu + G_a \omega^a_\mu + J\omega_\mu+ Db_\mu + K f_\mu  +  N m_\mu  \,.
\end{equation}
The curvature two-form associated with the time translation generator $H$ is given by
\begin{equation}
	R(H) = d\tau -2 b\wedge \tau \,.
\end{equation}
Imposing the curvature constraint $R(H)=0$ indicates $\tau \wedge d\tau=0$, which ensures that the foliation defined by $\tau_\mu$ is hypersurface orthogonal. This condition characterizes the twistless torsional Newton–Cartan (TTNC) geometry.

We now study the gauge transformation of the Schr\"odinger gauge field
$\mA_\mu$, as given in \eqref{eq:Gauge-transform-A}, under internal transformations parameterized by
\begin{equation}
	\Sigma = G_a \lambda_a + J\lambda +D\Lambda_D +K\Lambda_K + N \sigma
\end{equation}
Under this transformation, the component fields transform as follows:
\begin{align}
	\begin{split}
\bar{\delta} \tau_\mu   & = \mL_\xi \tau_\mu +2\Lambda_D \tau_\mu \\
\bar{\delta} e_\mu^a  & = \mL_\xi e_\mu^a +\lambda^a \tau_\mu +\Lambda_D e_\mu^a +\lambda \epsilon_{ab} e_\mu^b \\
\bar{\delta} \omega_\mu^a  & = \mL_\xi \omega_\mu^a + \partial_\mu \lambda^a + \lambda \epsilon_{ab} \omega_\mu^b - \omega_{\mu} \epsilon_{ab} \lambda^b +\lambda^a b_\mu -\Lambda_D \omega_\mu^a +\Lambda_K e_\mu^a \\
\bar{\delta} \omega_\mu  & = \mL_\xi \omega_\mu +\partial_\mu \lambda \\
\bar{\delta} b_\mu &= \mL_\xi \omega_\mu +\partial_\mu \Lambda_D +\Lambda_K \tau_\mu \\
\bar{\delta} f_\mu &= \mL_\xi f_\mu  + \partial_\mu\Lambda_K +2\Lambda_K b_\mu -2\Lambda_D f_\mu\\
\bar{\delta} m_\mu &= \mL_\xi m_\mu + \partial_\mu \sigma + \lambda^a e_\mu^a
	\end{split}
\end{align}
To define covariant derivatives that are invariant under local dilatations $D$ we compensate the $\Lambda_D$ transformations by modifying the connection as follows:
\begin{align}\label{eq:covariant-TTNC}
	\begin{split}
		\mathcal{D}_\mu \tau_\nu & = \partial_\mu \tau_\nu -\tilde{\Gamma}_{\mu \nu}^{\rho} \tau_\rho-2 b_\mu \tau_\nu \,, \\
		\mathcal{D}_\mu e_\nu^a & =\partial_\mu e_\nu^a   -\tilde{\Gamma}_{\mu \nu}^{\rho} e_\rho^a -\omega_\mu^a \tau_\nu -\omega_\mu \epsilon_{ab} e_\nu^b - b_\mu e_\nu^a  \,.
	\end{split}
\end{align}
Solving the vielbein postulates \eqref{eq:vielbein-postulate-TT} using these covariant derivatives yields the affine connection:
\begin{align}\label{eq:affine-TTNC-frame}
	\begin{split}
		\tilde{\Gamma}_{\mu\nu}^\rho = -\hat{v}^\rho (\partial_\mu \tau_\nu-2 b_\mu \tau_\nu) + e^\rho_a (\partial_\mu e_\nu^a -\omega_\mu^a \tau_\nu -\omega_\mu \epsilon_{ab} e_\nu^b -b_\mu e_\nu^a)
	\end{split}
\end{align}
Alternatively, we may express the connection in terms of the metric-like fields $\tau_\mu$ and $\bar{h}_{\mu\nu}$ as 
\begin{align}\label{eq:affine-TTNC-metric}
	\begin{split}
		\tilde{\Gamma}_{\mu\nu}^\rho = -\hat{v}^\rho (\partial_\mu \tau_\nu-2 b_\mu \tau_\nu) + \frac{1}{2} h^{\rho \sigma} \Big[(\partial_\mu-2b_\mu) \bar{h}_{\nu\sigma} +  (\partial_\nu-2b_\nu) \bar{h}_{\mu\sigma}- (\partial_\sigma-2b_\sigma) \bar{h}_{\mu\nu}\Big]
	\end{split}
\end{align}
Both expressions \eqref{eq:affine-TTNC-frame} and \eqref{eq:affine-TTNC-metric} are manifestly invariant under local Galilean boosts $G_a,J,D$ transformations. 
To ensure invariance under the central charge transformation $N$, one introduces a St\"uckelberg scalar $\chi$ and replaces $m_\mu$ in all boost-invariant quantities (such as in \eqref{eq:def-hatv-barh}) by the gauge-invariant combination: $M_\mu = m_\mu -\partial_\mu \chi$, as discussed in \cite{Bergshoeff:2014uea}.
Finally, both the affine connections \eqref{eq:affine-TTNC-frame} and \eqref{eq:affine-TTNC-metric} transform non-trivially under local special conformal transformations $K$ with the following transformation law:
\begin{equation}\label{eq:spec-trans-affine-TTNC}
	\bar{\delta} \tilde{\Gamma}_{\mu\nu}^\rho = -\Lambda_K (\delta_\mu^\rho \tau_\nu + \delta_\nu^\rho \tau_\mu)
\end{equation}
However, we emphasize that the transformation property \eqref{eq:spec-trans-affine-TTNC} does not imply that the two expressions for the affine connection—\eqref{eq:affine-TTNC-frame} and \eqref{eq:affine-TTNC-metric}—are identical.
Rather, they represent different but equivalent formulations of the same geometric structure. Importantly, the definitions of the Ricci tensor and other curvature invariants are independent of the specific choice of affine connection, provided the torsion constraints and metric compatibility conditions are satisfied.

\section{Chern-Simons theory by gauging so$(2,3)$}
\label{appendix:SO32}

The Chern-Simons theory by gauging so$(2,3)$ algebra is known to be the gravitational Chern-Simons term
\cite{Horne:1988jf,Merbis:2014vja,Kraus:2005zm}. 
In this section, we will rewrite the so$(2,3)$ algebra in terms of a  basis with Galilean boost manifest. 
This will be convenient to highlight the analogy between the Chern-Simons action in so$(2,3)$ algebra and the ones by su$(1,2)$ algebra \eqref{eq:CS-SU12}. 

We take the signature $\eta_{AB}=\text{Diag}(-,+,+)$ which denotes the tangent space of three dimensional manifold. 
The so$(2,3)$ generators are denoted as $P_A, J_{AB}, K_A,D$. 
The three dimensional invariant tensor $\epsilon_{ABC}$ could help to change the $J_{AB}$ generators into $J_A$ by the contraction $J_{AB} =  \epsilon_{ABC} J^{C}$.
This results in the algebra 
\begin{align}\label{eq:SO32-basis1}
	\begin{split}
		& [P_A,J_B]=\epsilon_{ABC} P^C, \qquad [K_A,J_B]=\epsilon_{ABC} K^C \\
		& [P_A,K_B]=\eta_{AB} D -\epsilon_{ABC} J^C, \qquad [J_A,J_B] =\epsilon_{ABC} J^C\\
		& [P_A,D]=P_A, \qquad [K_A, D]=-K_A\,.
	\end{split}
\end{align}
In this study, we would like to use an alternative basis of this algebra.
By splitting the Lorentzian into temporal and spatial parts following \eqref{eq:split-indices}, we make following redefinitions of the generators 
	\begin{equation}\label{eq:redef-generatorSo32}
		H= P_0, \quad K_0 = -K , \quad J_0 =J, \quad  J_a= \epsilon_{ab} G_b\,.
	\end{equation}
This basis makes the Galilean structure manifest in the relativistic algebra. 
Explicitly, the so$(2,3)$ algebra \eqref{eq:SO32-basis1} can then be reformulated as \footnote{
In this basis, the so$(2,2)$ algebra will be 
\begin{align}\label{eq:SO22-Galileanbasis-1}
	&[J,G_a] = \epsilon_{ab} G_b, \quad [J,P_a] =\epsilon_{ab} P_b, \quad [H, G_a] =P_a, \quad [H,P_a] = -G_a \\  \label{eq:SO22-Galileanbasis-2}
	& [P_a,P_b] = -[G_a,G_b] = \epsilon_{ab} J, \qquad  [P_a,G_b] = H \delta_{ab}\,.
\end{align}
}
\begin{align}\label{eq:SO32-basis2}
	\begin{split}
		& [J, P_a] =\epsilon_{ab} P_b, \quad [J,G_a] = \epsilon_{ab} G_b, \quad [J,K_a] = \epsilon_{ab} K_b \\
		& [H,D]= H, \quad [K,D]=-K, \quad [H,K]=D \\ 
		&  [H, G_a] = P_a, \quad [P_a ,K] = G_a, \quad [P_a,D]= P_a, \quad [K_a,D]=-K_a \\
		& [H,K_a] = G_a, \quad  [P_a, G_b]=\delta_{ab} H , \quad [P_a,K_b]= J \epsilon_{ab} +D \delta_{ab} \\
		& [G_a, G_b] = -J \epsilon_{ab}, \quad [K,G_a] = -K_a, \quad [K_a,G_b] = -K \delta_{ab}  \,.
	\end{split}
\end{align}
In this basis \eqref{eq:SO32-basis2}, we can write the bilinear product as \cite{Nappi:1993ie}
\begin{align}
	\begin{split}
		B(H,K) = B(P_a,K_a) = B(J,J) =-B(D,D)= -c_S, \quad B(G_a,G_b)=c_S\delta_{ab} \,.
	\end{split}
\end{align}

The gauge field is valued in this algebra as 
\begin{equation}
	\mathcal{A} = H E^0+ P_a E^a + J\Omega + G_a \Omega^a + K F+ K_a F^a+D B \,,
\end{equation}
where we have split the vielbein and spin connection $E^A, \Omega^A$ into temporal and spatial parts, respectively. The components of the strength tensors 
\begin{align}
	\begin{split}
		\mathcal{F}_{\mu\nu} &= H R_{\mu \nu } (H) + P_a R_{\mu\nu}^a(P) + G_a R_{\mu\nu}^a(G) + J R_{\mu\nu}(J) + D R_{\mu\nu}(D) + \\
		& + K_a R_{\mu\nu}^a (K) + K R_{\mu\nu}(K)
	\end{split}
\end{align}
are respectively
\begin{align}
	\begin{split}
		R(H) &= dE^0 +E^0 \wedge B+ E^a \wedge \Omega^a \\
		R^a (P) &= dE^a + E^0 \wedge \Omega^a+ \epsilon_{ab} E^b\wedge \Omega + E^a \wedge B \\
		R^a (G) &= d\Omega^a+ E^0 \wedge F^a +E^a \wedge F - \epsilon_{ab} \Omega \wedge \Omega^b \\
		R (J) &= d\Omega+ \epsilon_{ab}E^a \wedge F^b -\Omega^1 \wedge \Omega^2 \\
		R(D) &= dB + E^a \wedge F^a+ E^0 \wedge F \\
		R(K) &= dF + \Omega^a \wedge F^a -F \wedge B \\
		R^a(K) &= dF^a  - \epsilon_{ab}\Omega \wedge F^b + \Omega^a \wedge F -F^a \wedge B \,.
	\end{split}
\end{align}
From the perspective of Newton-Cartan gravity, the curvature tensor $R(H)=0$ indicates the existence of a general torsion tensor, as $E^0 \wedge d E^0 \neq 0$.
The transformations under the internal gauge symmetry
\begin{equation}
	\Sigma = G_a \Lambda_a + J \Lambda + D \Lambda_D +  K \Lambda_K +\Lambda^a_K K_a
\end{equation}
 of each components are then 
\begin{align}\label{eq:SO32-transform-fields}
	\begin{split}
		\delta E^0 &= \Lambda_D E^0 +E^a \Lambda^a \\
		\delta E^a &= \Lambda^a E^0 +E^a \Lambda_D +\epsilon_{ab} E^b \Lambda \\
		\delta \Omega^a &= d\Lambda^a +E^0 \Lambda_K^a + E^a \Lambda_K +\epsilon_{ab} \Omega^b \Lambda- \Omega \epsilon_{ab} \Lambda^b \\
		d\Omega &= d\Lambda + \epsilon_{ab} E^a \Lambda_K^b+ \epsilon_{ab} \Lambda^a \Omega^b \\
		dB &= d\Lambda_D + E^0 \Lambda_K + E^a \Lambda_K^a \\
		dF &= d\Lambda_K +\omega^a\Lambda_K^a + B \Lambda_K -F \Lambda_D -F^a \Lambda^a \\
		dF^a &= d\Lambda_K^a +\Omega^a \Lambda_K- \Omega \Lambda_K^b \epsilon_{ab} + B \Lambda_K^a -F \Lambda^a+\Lambda \epsilon_{ab} F^b -F^a \Lambda_D  \,.
	\end{split}
\end{align}
With the orthogonal condition, the  inverse vielbein are
\begin{align}\label{eq:inverse-transform}
	\begin{split}
		\delta E^\mu_a = -\Lambda_D E^\mu_a +\epsilon_{ab} E_b^\mu \Lambda +E^\mu_0 \Lambda^a, \qquad \delta E^\mu_0 = -E_0^\mu\Lambda_D +E^\mu_a \Lambda^a\,.
	\end{split}
\end{align}

The Lagrangian of the so$(2,3)$ Chern-Simons theory can then be computed as 
\begin{align}\label{eq:Lag-SO23}
	\begin{split}
		\frac{\mathcal{L}}{c} &= -2F \wedge R(H) -2 F^a \wedge R^a (P) \\
		& -\Omega \wedge d\Omega + \Omega^a \wedge d \Omega^a +2\Omega \wedge \Omega^1 \wedge \Omega^2+ B \wedge dB \,.
	\end{split}
\end{align}
The gauge fields $F,F^a$ are associated with special conformal transformations $K,K^a$, acting as the Lagrange multiplers to impose the curvature constraints 
\begin{equation}
	R(H)= R^a(P)=0\,.
\end{equation}
Then the remaining action \eqref{eq:Lag-SO23} can be shown to be the gravitational Chern-Simons term 
\begin{equation}\label{eq:threeGamma}
	S_{gCS} = \int d^3 x \epsilon^{\mu\nu\rho}\Gamma_{\mu\sigma}^{\lambda}\left(\partial_{\nu}\Gamma_{\rho\lambda}^{\sigma}+\frac{2}{3}\Gamma_{\nu\tau}^{\sigma}\Gamma_{\rho\lambda}^{\tau}\right)\,. 
\end{equation}
up to the boundary terms \cite{Merbis:2014vja,Kraus:2005zm}. 
Here the affine connection is determined by vielbein postulates:
\begin{align}\label{eq:vielbein-postu}
	\begin{split}
	\nabla_\mu E_\nu^0 &= \partial_\mu E_\nu^0  -\Omega_\mu^a E_\nu^a -\Gamma_{\mu\nu}^\sigma E_\sigma^0 \\
	\nabla_{\mu} E_\nu^a &= \partial_\mu E_\nu^a  -\Omega_\mu \epsilon_{ab} E_\nu^b - \Omega_\mu^a E_\nu^0 -\Gamma_{\mu\nu}^\sigma E_\sigma^a \,.
	\end{split}
\end{align}
While the $\Gamma_{\mu\nu}^\sigma$ is not invariant under the local dilatation transformation,  
this invariance can be restored by replacing $\partial$ with $\partial- B$, which fixes the affine connection to be
\begin{align}\label{eq:affine-SO32}
	\begin{split}
		\tilde{\Gamma}_{\mu\nu}^\sigma &= -E_0^\sigma (\partial_\mu -B_\mu) E_\nu^0 + E_a^\sigma (\partial_\mu -B_\mu) E_\nu^a \\
		&+ \Omega_\mu^a(E^\sigma_0 E_\nu^a -E^\sigma_a E_\nu^0) -\Omega_\mu (E_a^\sigma \epsilon_{ab} E_\nu^b) \,.
	\end{split}
\end{align}
The affine connection \eqref{eq:affine-SO32} is invariant under $\Lambda_D,\Lambda_a,\Lambda$ transformations: 
\begin{equation}\label{eq:transform-affine-So32}
	\delta \tilde{\Gamma}_{\mu\nu}^\sigma = \Lambda_K \left(
	-E_\mu^0 \delta_\nu^\sigma- \delta_\mu^\sigma E_\nu^0
	+g_{\mu\nu} E^\sigma_0\right) + \left(
	-E_\mu^a \delta_\nu^\sigma-E_\nu^a \delta_{\mu}^\sigma + E_a^\sigma g_{\mu\nu}
	\right) \Lambda_K^a \,.
\end{equation}
We can also use \eqref{eq:transform-affine-So32} to rewrite the gravitational Chern-Simons term.
The final gravitational action is equivalent to the one determined by \eqref{eq:threeGamma}.

Since this is a Lorentzian theory, the invariant metric-like fields are given explicitly by the Lorentzian metric
\begin{equation}
	g_{\mu\nu} = - E_\mu^0 E_\nu^0 +E_\mu^a E_\nu^b \delta_{ab}  \,.
\end{equation}
The affine connection $\tilde{\Gamma}_{\mu\nu}^\rho$ in terms of metric-like fields is:
\begin{align}\label{eq:affineSo32-metric}
	\begin{split}
		\tilde{\Gamma}_{\mu\nu}^\sigma &= \frac{1}{2} g^{\rho \sigma} \left[
		(\partial_\mu-2B_\mu) g_{\rho \nu} + (\partial_\nu-2B_\nu) B_{\mu\rho} -(\partial_\rho -2B_\rho) g_{\mu\nu}
		\right] \,,
	\end{split}
\end{align}
which can be shown to follow the same transformation as \eqref{eq:transform-affine-So32}. 
The physical theory is independent of the chosen affine connection.

\section{su$(1,3)$ algebra} \label{appendix:SU13}
The su$(1,3)\oplus$u$(1)$ algebra arises from the null reduction of the 6d CFT \cite{Lambert:2020zdc, Lambert:2019jwi, Lambert:2020scy, Lambert:2024uue}.
Mathematically, the complexified Lie algebras of su$(1,3)$ and so$(2,4)$ are isomorphic. 
Therefore, it can also be treated as the conformal algebra of four dimensional conformal field theory. 
In this work, we  will emphasize its use as the conformal algebra for five dimensional non-Lorentzian spacetime. 
The algebra consists of time translation $H$, spatial translation $P_i$, dilatation $D$, Galilean boost $G_i$, rotation triplet $J^\alpha$ parametrizing SU$(2)$=SO$(3)$, rotation singlet $B$ and
one special conformal transformation $K$, where $i=1,2,3,4$.

The non-vanishing commutators of su$(1,3)\oplus$u$(1)$ algebra include \cite{Lambert:2021nol}: 
\begin{align}\label{eq:SU13algebra}
	\begin{split}
		&[P_i,P_j] = \Omega_{ij} H, \quad [G_i,G_j] =\Omega_{ij} K,\quad [J^\alpha,J^\beta]=f^{\alpha\beta}{}_\gamma J^\gamma\\
		&[B,G_i]=\Omega_{ij}P_j,\quad[B,P_i]=\Omega_{ij}G_j,\quad[J^\alpha,P_i]=-L^{\alpha}_{ij}P_j\\
		&[D,H]=-2H,\quad[D,K]=2K,\quad[J^\alpha,G_i]=-L^\alpha_{ij}G_j\\
		&[H,G_i]=P_i,\quad[K,P_i]=-G_i,\quad[K,H]=-D\\
		&[D,P_i]=-P_i,\quad[D,G_i]=G_i, \quad [P_i,G_j]=\frac{1}{2}\Omega_{ij}D-\delta_{ij}B-\beta^\alpha_{ij}J^\alpha + \delta_{ij} N \,,
	\end{split}
\end{align}
where $\Omega_{ij}$ is a constant anti-symmetric matrix that satisfies $\Omega_{ij}\Omega_{jk}=-\delta_{ik}$. 
Its explicit form is:
\begin{equation}\label{eq:MatrixOmega-5d}
	\Omega_{ij} = \left(
	\begin{array}{cc}
		0	& \mathbbm{1}_{2} \\
		-\mathbbm{1}_{2}	&  0
	\end{array}
	\right)\,.
\end{equation}
The constants $f^{\alpha\beta}{}_{\gamma}$ are the structure constant of su$(2)$ while the generators $L^{\alpha}_{ij}$ satisfy $[L^\alpha,L^\beta]_{ij}=f^{\alpha\beta}{}_{\gamma}L^{\gamma}_{ij}$.
Finally, $\beta^{\alpha}_{ij}$ are constants defined as follows:
\begin{equation}
	\beta^{\alpha}_{ij}L^{\alpha}_{kl}=\frac{1}{2}(\delta_{jk}\Omega_{il}+\delta_{ik}\Omega_{jl}-\delta_{jl}\Omega_{ik}-\delta_{il}\Omega_{jk}-\delta_{ij}\Omega_{kl}) \,.
\end{equation}
A few properties of the Newton-Cartan gravity by gauging su$(1,3)\oplus$u$(1)$ algebra \eqref{eq:SU13algebra} can be read from the algebra. 
\begin{itemize}
	\item First of all, the non-vanishing commutators of $[P_i,P_j] = \Omega_{ij} H$ include
\begin{equation}
	[P_1,P_3]= [P_2,P_4] = H \,.
\end{equation} 
This fixes the curvature tensor related to torsion as
\begin{equation}
	R(H) = d\tau -2b\wedge \tau +e^1 \wedge e^3+e^2 \wedge e^4=0 \,.
\end{equation}
which indicates that the geometry is completely a torsional Newton-Cartan type, as $\tau \wedge d\tau \neq 0$.

\item The non-trivial dependence of $D,J^\alpha$ generators in the  commutator $[P_i,G_j]$ indicates that the corresponding gauge fields $b_\mu$ and the spin connection $\Omega^\alpha$ are non-invariant under the Galilean boost transformation. 
The corresponding affine connection constructed from the vielbein postulate therefore can be made dilatation invariant but not boost invariant, similar to the example studied in three dimensions \eqref{eq:affine-with-Bvari-gauge}. 
This is a typical feature of the torsional Newton-Cartan gravity  built up by gauging su$(1,n)$ type algebra and is distinguished from the structures discussed in \cite{Bergshoeff:2014uea}. 
\end{itemize}

\bibliographystyle{JHEP}
\bibliography{bib-bh}
 
\end{document}